\documentclass[graybox]{svmult}
\usepackage{amssymb,amsmath,enumerate}
\usepackage{multirow}
\usepackage{bm}
\usepackage{epsfig}
\usepackage{multicol}        % used for the two-column index
\usepackage{url}
\usepackage{authblk}
\usepackage{bbold}
\usepackage{mathptmx}       % selects Times Roman as basic font
\usepackage{helvet}         % selects Helvetica as sans-serif font
\usepackage{courier}        % selects Courier as typewriter font
\usepackage{type1cm}        % activate if the above 3 fonts are
\usepackage{makeidx}         % allows index generation
\usepackage{graphicx}        % standard LaTeX graphics tool
\usepackage{multicol}        % used for the two-column index
\usepackage[bottom]{footmisc}% places footnotes at page bottom
\makeindex             % used for the subject index
\graphicspath{{./Figures/}}
\begin{document}
\title*{Complex market dynamics in the light of random matrix theory}
\author{Hirdesh K. Pharasi,  Kiran Sharma, Anirban Chakraborti and Thomas H. Seligman}
\institute{Hirdesh K. Pharasi \at Instituto de Ciencias F\'{i}sicas, Universidad Nacional Aut\'{o}noma de M\'{e}xico, Cuernavaca-62210, M\'{e}xico, \email{hirdeshpharasi@gmail.com}
\and  Kiran Sharma \at School of Computational and Integrative Sciences, Jawaharlal Nehru University, New Delhi-110067, India, \email{kiransharma1187@gmail.com }
\and  Anirban Chakraborti \at School of Computational and Integrative Sciences, Jawaharlal Nehru University, New Delhi-110067, India, \email{anirban@jnu.ac.in}
\and  Thomas H. Seligman \at Instituto de Ciencias F\'{i}sicas, Universidad Nacional Aut\'{o}noma de M\'{e}xico, Cuernavaca-62210, M\'{e}xico and Centro Internacional de Ciencias, Cuernavaca-62210, M\'{e}xico \email{seligman@icf.unam.mx}}
\maketitle

\abstract{We present a brief overview of random matrix theory (RMT) with the objectives of highlighting the computational results and applications in financial markets as complex systems. An oft-encountered problem in computational finance is the choice of an appropriate epoch over which the empirical cross-correlation return matrix is computed. A long epoch would smoothen the fluctuations in the return time series and suffers from non-stationarity, whereas a short epoch results in noisy fluctuations in the return time series and the correlation matrices turn out to be highly singular. An effective method to tackle this issue is the use of the power mapping, where a non-linear distortion is applied to a short epoch correlation matrix. The value of distortion parameter controls the noise-suppression. The distortion also removes the degeneracy of zero eigenvalues. Depending on the correlation structures, interesting properties of the eigenvalue spectra are found. We simulate different correlated Wishart matrices to compare the results with empirical return matrices computed using the S\&P 500 (USA) market data for the period 1985-2016. We also briefly review two recent applications of RMT in financial stock markets: (i) Identification of ``market states'' and long-term precursor to a critical state; (ii) Characterization of catastrophic instabilities (market crashes).}
%%>>>>>>>>>>>>>>>>>>>>>>>>>>>>>>>>>>>>>>>>>>>>>>>>>>>>>>>>>>>
\section{Introduction}

With the advent of the ``Big Data'' era \cite{bigdata_2014,bigdata_2015}, large data sets have become ubiquitous in numerous fields -- image analysis, genomics, epidemiology, engineering, social media, finance, etc., for which we need new statistical and analytical methods \cite{Bouchaud_2003,Chakraborti_2011a,Chakraborti_2011b,Mantegna_2007,Sinha_2010}.
Empirical correlation matrices are of primal importance in big data analyses, since various statistical methods strongly rely on the validity of such matrices in order to isolate meaningful information contained in the ``observational'' signals or time series \cite{bendat_1980}. Often the time series are of finite lengths, which can lead to spurious correlations and make it difficult to extract the signal from noise \cite{Guhr_2003,Schafer_2010}. Hence, it is very important to understand quantitative effects of finite-size time series in determination of empirical correlations \cite{Chakraborti_2018,Guhr_2003,Schafer_2010,Schafer_2013}.

Random matrix theory (RMT) tries to describe statistics of eigenvalues of random matrices, often in the limit of large dimensions. The subject came up first in a celebrated paper of J. Wishart~\cite{wishart_1928} in 1929 where he proposed that the correlation matrix of white noise time series was an adequate prior for correlation matrices. E. Cartan proposed the classical random matrix ensembles in an important but little known paper~\cite{cartan_1935}. After that there was increasing interest in the subject among which it is important to mention work by L.G. Hua, who published the first monographs on the subject in 1952; an English translation is available~\cite{hua_1963}. 

Wigner introduced RMT to physics, based on the assumption that the interactions between the nuclear constituents were so complex that they could be modeled as random fluctuations in the framework of his R-matrix scattering theory~\cite{wigner_1951}. This culminated in the presentation of the Hamiltonian $\hat{H}$ as a large random matrix, such that the energy levels of the nuclear system could be approximated by the eigenvalues of this matrix, and indeed the spacings between the energy levels of nuclei could be modeled by the spacing of eigenvalues of the matrix~\cite{wigner_1958, wigner_1967}. The use of RMT has spread over many fields from molecular physics~\cite{leviandier_1986} to quantum chromodynamics~\cite{shuryak_1993}. Lately, RMT has become a popular tool for investigating the dynamics of financial markets using cross-correlations of empirical return time series \cite{plerou_1999,utsugi_2004}.

In this chapter, we present recent techniques of random matrix theory (RMT) mainly focused on computational results and applications of correlations in financial markets viewed as complex systems \cite{Yaneer_2002,Gellmann_1995,utsugi_2004,Vemuri_1978}. A central problem that often arises in computational finance  is the choice of the epoch size over which the empirical cross-correlation return matrix needs to be computed. A very long epoch would smoothen the fluctuations in return time series and also the time series suffers  from the problem of non-stationarity \cite{mikosch_2004}, whereas a short-time epoch would result in noisy fluctuations in return time series and the correlation matrix turns out to be highly singular (with many zero eigenvalues) \cite{Chakraborti_2018}. Among others, an effective method to tackle this issue has been the use of the power mapping \cite{Chakraborti_2018,Guhr_2003,Schafer_2010,Schafer_2013}, where a non-linear distortion is applied to a short epoch correlation matrix. Here, we demonstrate how the value of distortion parameter controls the noise-suppression. It also removes the degeneracy of the zero eigenvalues (which for very small values of the distortion parameter leads to a well separated ``emerging spectum" near zero). Depending on the correlation structures, interesting properties of the eigenvalue spectra are found. Correlation matrices constructed from white noise were introduced by Wishart and their eigenvalue spectrum gets a shape of Mar\u{c}enko-Pastur distribution \cite{Marcenko_1967}; there are significant deviations when a correlation structure is introduced \cite{Chakraborti_2007}. We simulate different correlated Wishart matrices \cite{Mehta_2004,wishart_1928} to compare the results with empirical return matrices computed using S\&P 500 (USA) market data for the period 1985-2016 \cite{Chakraborti_2018}. We also briefly review two recent applications of RMT in financial stock markets: (i) Identification of ``market states'' and long-term precursor to a critical state \cite{Pharasi_2018}; (ii) Characterization of catastrophic instabilities (market crashes) \cite{Chakraborti_2018}.

This paper is described as follows.  Section 2 discusses the data description, methodology and results in details. 
Section 3 contains applications of RMT in financial markets. Finally, section 4 contains concluding remarks.

%%>>>>>>>>>>>>>>>>>>>>>>>>>>>>>>>>>>>>>>>>>>>>>>>>>>>>>>>>>>>>>>>>>>>>>>>>>>>>>>>>>>>>>>>>>>
\section{Data Description, Methodology and Results}

\subsection{Data description}
\label{Sec:Materials}
We have used the database of Yahoo finance \cite{Yahoo_finance}, for the time series of adjusted closure prices for S\&P 500 (USA) market, for the period 02/01/1985 to 30/12/2016 ($T=8068$ days); number of stocks $N=194$, where we have included the stocks that are present in the index for the entire duration. The sectoral abbreviations are given in Table~\ref{Table:sectoral_index_sp500}.

%%=====================================================
\begin{table}[h]
\centering
\caption{Abbreviations of ten different sectors for S\&P 500 index}
\label{Table:sectoral_index_sp500}
\begin{tabular}{|l|l|l|l|}
\hline
\textbf{Labels}  & \textbf{Sectors}        & \textbf{Labels} & \textbf{Sectors}  \\ \hline
\textbf{CD}      & Consumer Discretionary  & \textbf{ID}     & Industrials       \\ \hline
\textbf{CS}      & Consumer Staples        & \textbf{IT}     & Information Technology\\ \hline
\textbf{HC}      & Health Care 	           & \textbf{MT}     & Materials              \\ \hline
\textbf{EG}      & Energy                  & \textbf{TC}     & Technology              \\ \hline
\textbf{FN}      & Financials              & \textbf{UT}     & Utilities             \\ \hline
\end{tabular}
\end{table}
%%=====================================================
\subsection{Methodology and Results}

Correlations between different financial assets play fundamental roles in the analyses of portfolio management, risk management, investment strategies, etc. However, one only has finite time series of the assets prices; hence, one cannot calculate the exact correlation among assets, but only an approximation. 
%Furthermore due to the finite size of time series,  noise is present. 
The quality of the estimation of the true cross-correlation matrix strongly depends on the ratio between the length of the financial price time series $T$ and the number of assets $N$. The larger the ratio  $Q = T/N$, the better the estimation is; though for practical limitations, the ratio can be even smaller than unity. However, such correlation matrices are often too noisy, and thus need to be filtered from noise. To build the correlation matrices, we first calculate the return $r_i$ from the daily price $P_i$ of stocks $i=1, . . . ,N$,  at time $t$ (trading day):
\begin{equation}
r_i(t)=\ln P_i(t)- \ln P_i(t-1),
\label{eq_return}
\end{equation}
where $P_i(t)$ denotes the price of stock $i$ at time $t$. Since different stocks have varying levels of volatility, we define the equal-time Pearson cross-correlation coefficient as 
\begin{equation}
C_{ij}(\tau) = \frac{\langle r_i r_j \rangle - \langle r_i \rangle \langle r_j \rangle}{\sigma_i\sigma_j},
\label{eq_corr}
\end{equation}
where $\langle \dots \rangle $ denotes the time average and $\sigma_k$ denotes the standard deviation of the return time series $r_k$, $k=1, \dots, N$, computed over an epoch of $M$ trading days ending on day $\tau$. The elements $C_{ij}$ are restricted to the domain $ -1\leq C_{ij}\leq1$, where $C_{ij}=1$ corresponds to perfect correlations, $C_{ij}=-1$ to perfect anti-correlations, and $C_{ij}=0$ to uncorrelated pairs of stocks. The difficulties in analyzing the significance and meaning of the empirical cross-correlation coefficients $C_{ij}$ are due to several reasons, which include the following:
\begin{enumerate}
\item Market conditions change with time and the cross-correlations that exist between any pair of stocks may not be stationary if an epoch chosen is too long.
\item Too short epoch for estimation of cross-correlations introduces ``noise'', i.e., fluctuations.
\end{enumerate}
For these reasons, the empirical cross-correlation matrix $\boldsymbol C (\tau)$ often contains ``random'' contributions plus a part that is not a result of randomness \cite{Plerou_2002, Pandey_2010}.
Hence, the eigenvalue statistics of $\boldsymbol C (\tau)$ are often compared against those of a large random correlation matrix -- a correlation matrix constructed from mutually uncorrelated time series (white noise) known as Wishart matrix.

We first reproduce the basic results of RMT, e.g., the Mar\u{c}enko-Pastur distribution, or Mar\u{c}enko-Pastur law, which describes the asymptotic behavior of eigenvalues of square random matrices \cite{Marcenko_1967}. Then, we present a study of time evolution of the empirical cross-correlation structures of return matrices for $N$ stocks and the eigenvalues spectra over different time epochs, and try to extract some new properties or information about the financial market \cite{Chakraborti_2018,Pharasi_2018}.

\subsubsection{Wishart and Correlated Wishart Ensembles}
Let us construct a large random matrix $\boldsymbol B$ arising from $N$ random time series each of length $T$, where the entries of a time series are real independent random variables drawn from a standard Gaussian distribution with zero mean and variance $\sigma^2$, such that the resulting matrix $\boldsymbol B$ is $N \times T$. Then the Wishart matrix can be constructed as
\begin{equation}
{\boldsymbol W}=\frac{1}{T}{\boldsymbol  B}{\boldsymbol B}'.
\label{eq_WC}
\end{equation}
In RMT, the ensemble of Wishart matrices is known as the \textit{Wishart orthogonal ensemble}.
In the context of a time series, $\boldsymbol{W}$ may
be interpreted as the \textit{covariance} matrix, calculated over $N$ stochastic time series, each with $T$ statistically
independent variables. This implies that on average, $\boldsymbol{W}$ does not have cross-correlations.

A correlated Wishart matrix can be constructed as
\begin{equation}
{\boldsymbol W}=\frac{1}{T}{\boldsymbol  G}{\boldsymbol G}',
\label{eq_CWC}
\end{equation}
where $\boldsymbol G= {\boldsymbol \zeta}^{1/2} \boldsymbol B$, is a $ N \times T$ matrix; ${\boldsymbol G}'$ is the $T \times N$ transpose matrix of $\boldsymbol G$, and the $N \times N$ positive definite symmetric matrix $\zeta$ controls the actual correlations. If $\boldsymbol{\zeta}$ is a diagonal matrix with the diagonal entries as unity and off-diagonal entries as zero (i.e., $\zeta=\mathbb{1}$, the identity matrix), then the resulting matrix $\boldsymbol{W}$ reduces to one of the former \textit{Wishart orthogonal ensemble}. If the  diagonal entries of $\boldsymbol{\zeta}$ are unity and off-diagonal elements are non-zero and real, then the resulting matrices form the \textit{correlated Wishart orthogonal ensemble}. For simplicity, in this chapter, we have generated and used $\boldsymbol{\zeta}$ for which all the off-diagonal elements are same (equal to a constant $U$, which lies between zero and unity).

The spectrum of eigenvalues for the Wishart orthogonal ensemble can be calculated analytically. For the limit $N \rightarrow \infty $ and $T \rightarrow \infty$, with $Q = T /N$  fixed (and bigger than unity), the probability density function of the eigenvalues is given by the Mar\u{c}enko-Pastur distribution:
\begin{equation}
\bar{\rho}(\lambda)=\frac{Q}{2\pi \sigma^2}\frac{\sqrt{(\lambda_{max}-\lambda)(\lambda-\lambda_{min})}}{\lambda},
\label{eq_RM}
\end{equation}
where $\sigma^2$ is the variance of the elements of $\boldsymbol G$,  while $\lambda_{min}$ and $\lambda_{max}$ satisfy the relation:
\begin{equation}
\lambda_{min}^{max}= \sigma^2 \left(1\pm \frac{1}{\sqrt{Q}}\right)^2 .
\label{eq_lambda}
\end{equation}

For $Q \leq 1$, positive semi-definite matrices $\boldsymbol{W}$, the density $\bar{\rho}(\lambda)$ in the above Eq.~\ref{eq_RM} is normalized to $Q$ and not to unity. Therefore, taking into account the $(N-T)$ zeros, we have

\begin{equation}
\bar{\rho}(\lambda)=\frac{Q}{2\pi \sigma^2}\frac{\sqrt{(\lambda_{max}-\lambda)(\lambda-\lambda_{min})}}{\lambda} + (1-Q)\delta(\lambda).
\label{eq_RM1}
\end{equation}

%\section{Results and Discussion}\label{Sec:Results}

\begin{figure}[!t]
\centering
\includegraphics[width=0.33\linewidth]{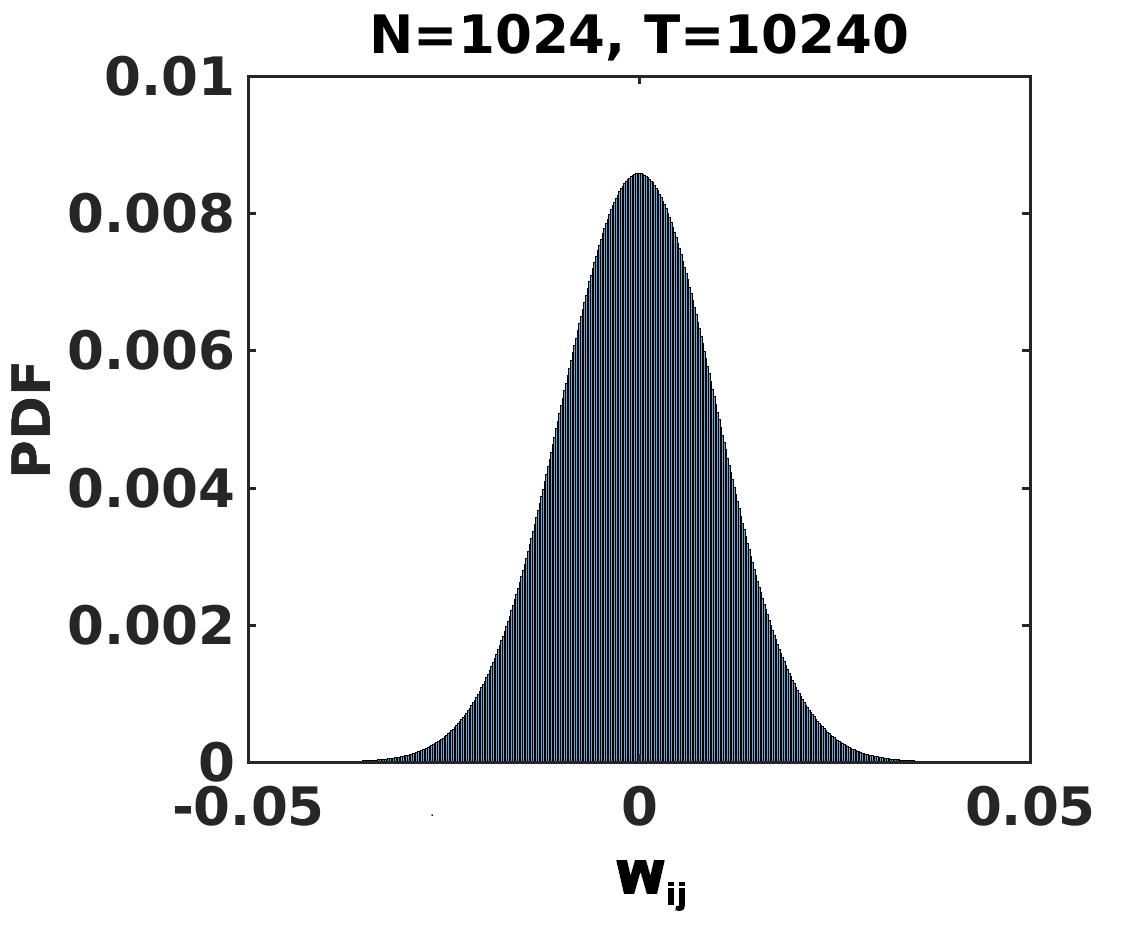}\llap{\parbox[b]{1.35in}{(\textbf{a})\\\rule{0ex}{1.2in}}}\includegraphics[width=0.33\linewidth]{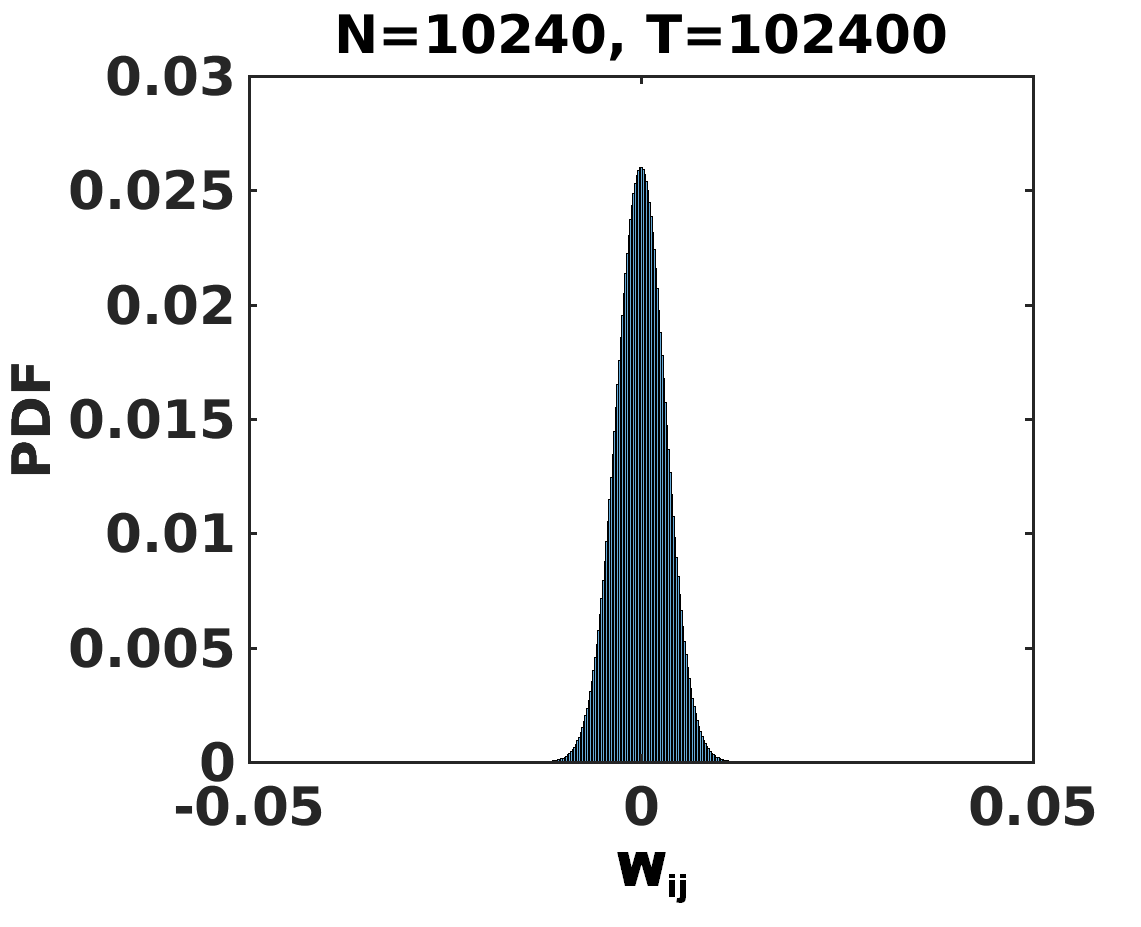}\llap{\parbox[b]{1.35in}{(\textbf{b})\\\rule{0ex}{1.2in}}}\includegraphics[width=0.33\linewidth]{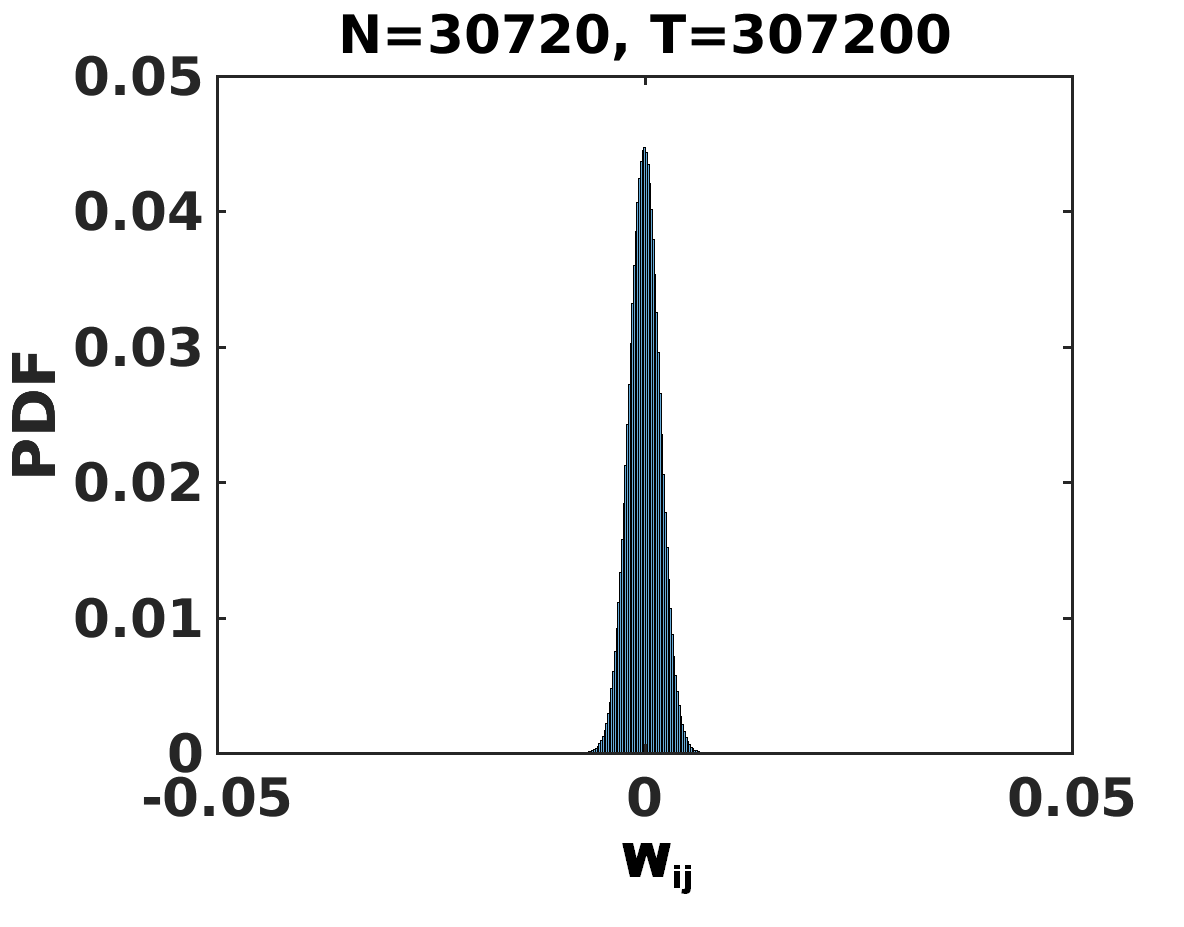}\llap{\parbox[b]{1.35in}{(\textbf{c})\\\rule{0ex}{1.2in}}}\\
\includegraphics[width=0.33\linewidth]{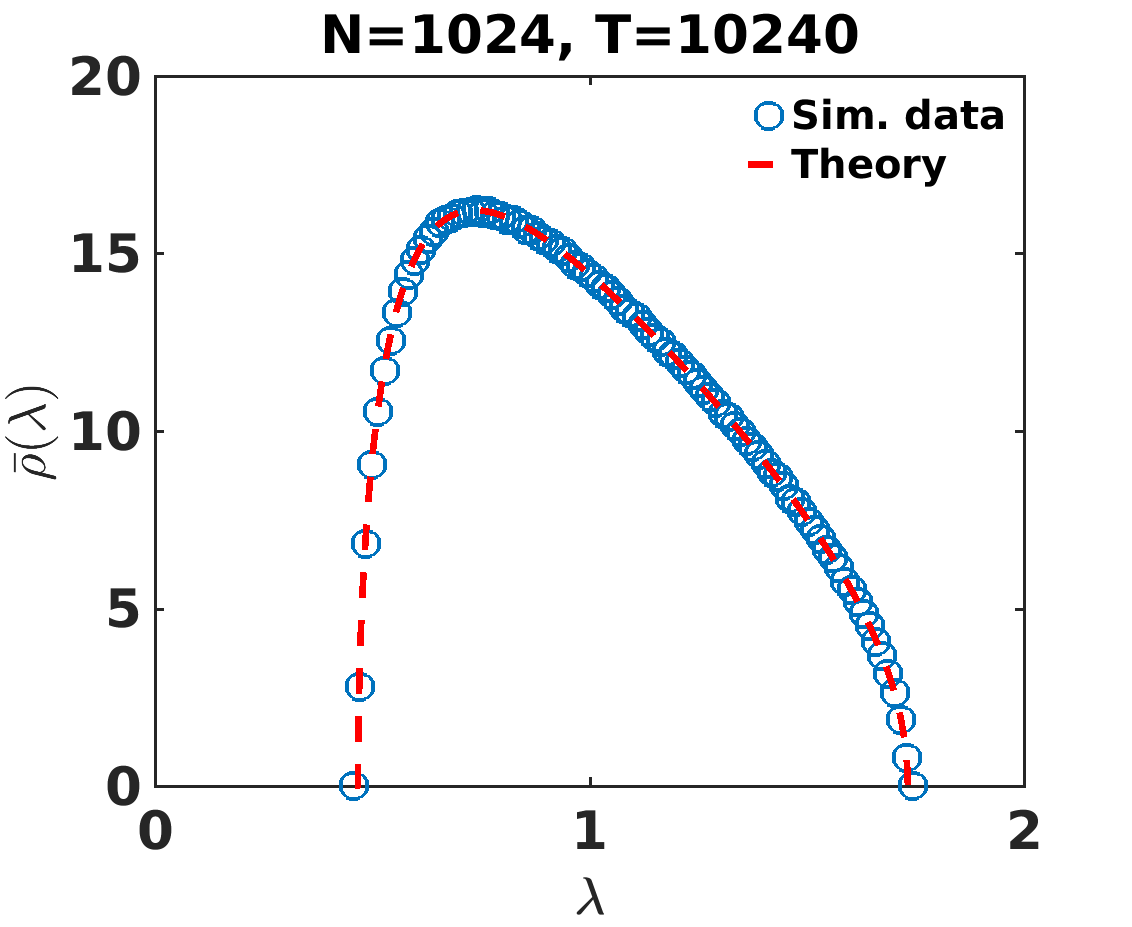}\llap{\parbox[b]{1.35in}{(\textbf{d})\\\rule{0ex}{1.25in}}}\includegraphics[width=0.33\linewidth]{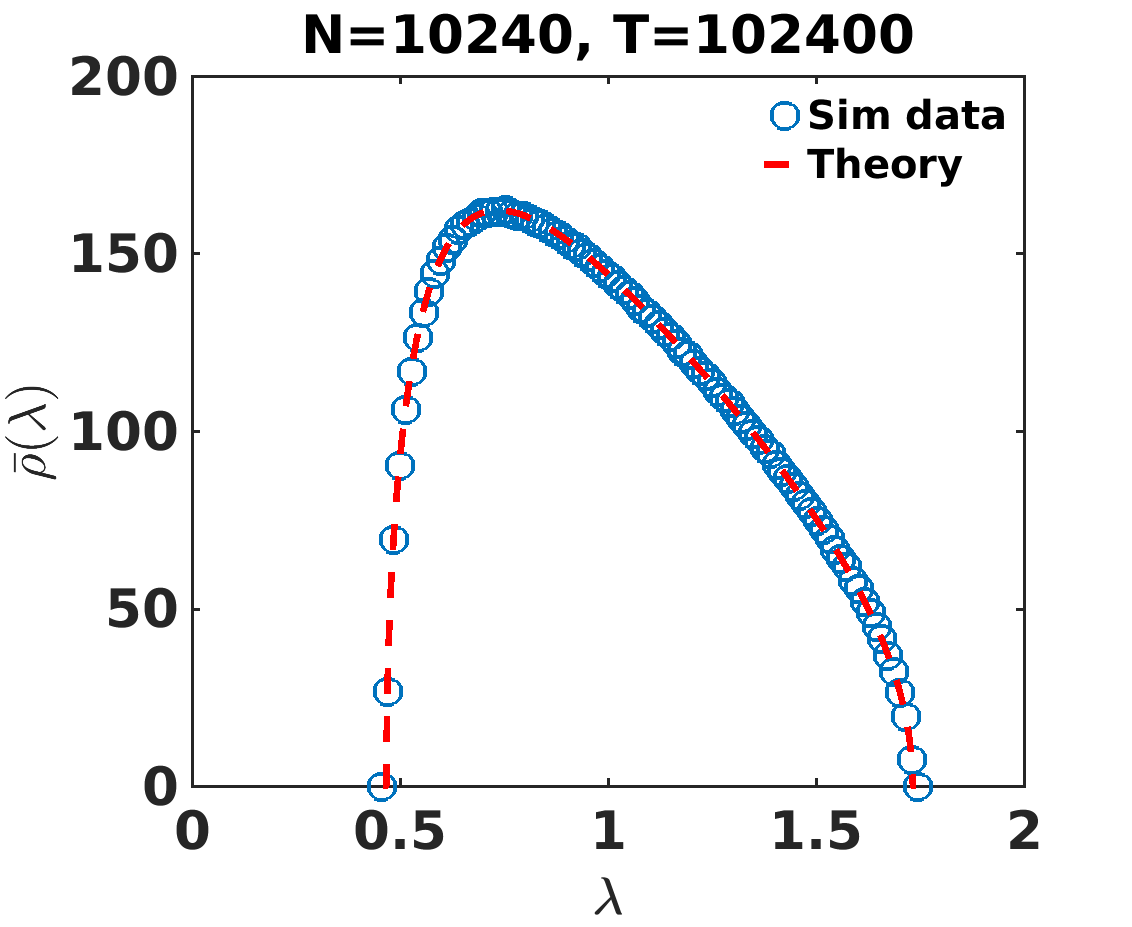}\llap{\parbox[b]{1.35in}{(\textbf{e})\\\rule{0ex}{1.25in}}}\includegraphics[width=0.33\linewidth]{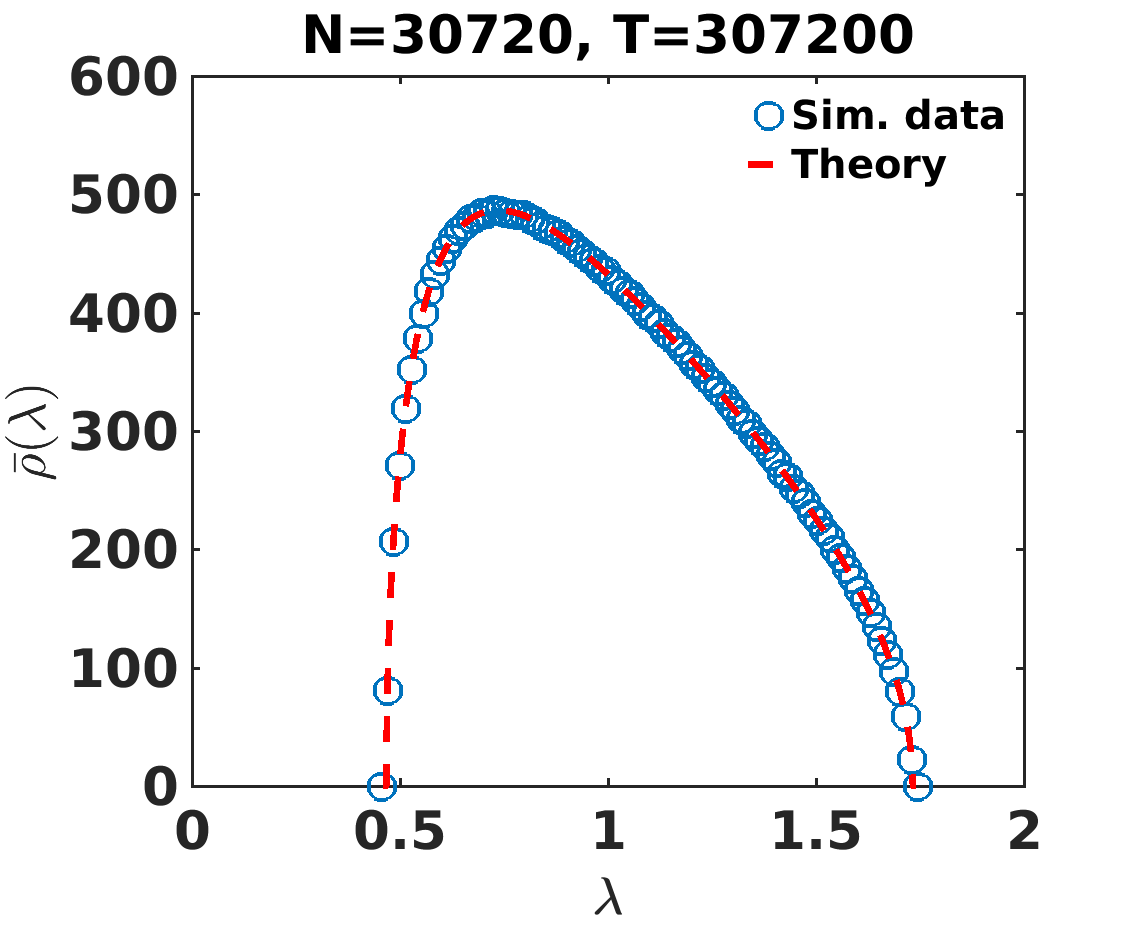}\llap{\parbox[b]{1.35in}{(\textbf{f})\\\rule{0ex}{1.25in}}}\\
\caption{(a)-(c) show the effect of finite size on true correlations with the dimensions of $\boldsymbol B$ ($N=$ finite, $T=$ finite and $Q(=T/N)=10$). The probability distribution of elements $(W_{ij})$ of the Wishart ensemble of size, constructed from $N$ time series, each with real independent Gaussian random variables of length $T$ with zero mean and variance $\sigma^2$. The variance of the distribution of $W_{ij}$ decreases with the increase of $N$ and $T$ and reduces to zero for $N\rightarrow \infty$ and $T\rightarrow\infty$ with $\frac{T}{N}=$ finite. (d)-(f) show the density of eigenvalues $\bar{\rho}(\lambda)$ of Wishart ensemble, which are numerically fitted with the Mar\u{c}enko-Pastur distributions~\cite{Marcenko_1967} (red dash lines) for all $N$ and $T$. The numerical values of $\lambda_{max}=1.732$ and $\lambda_{min}=0.468$ of the spectra also match exactly with the results theoretically calculated from Eq.~\ref{eq_lambda}. Numerical results for the probability distributions of the elements $(W_{ij})$ and densities of the eigenvalues $(\bar{\rho}(\lambda))$ have been generated using averages up to 200 ensembles.}
\label{fig_Fig1}
\end{figure}

First, we have generated a Wishart matrix $\boldsymbol W$ (with $\zeta=\mathbb{1})$ of size $N \times N$ constructed from $N$ time series of real independent Gaussian variables, each of finite length $T$, zero mean and unit variance ($\sigma^2 = 1$).
Fig.~\ref{fig_Fig1} shows the effect of finite sizes of the sets of parameters $N$ and $T$ on the probability distributions of the elements $W_{ij}$ of the Wishart ensemble and the corresponding eigenvalue spectra. Fig.~\ref{fig_Fig1} (a) shows the probability distribution of the elements of the Wishart matrix of dimensions, where $N=1024$ and $T=10240$. Fig.~\ref{fig_Fig1} (d) shows the corresponding density of eigenvalues $\bar{\rho}(\lambda)$, which takes the shape of the theoretical Mar\u{c}enko-Pastur distribution (red dashed line)~\cite{Marcenko_1967}. Similarly, Figs.~\ref{fig_Fig1} (b) and (c) show the respective probability distributions of the elements of Wishart matrices generated using the sets of parameters $N=10240$ and $T=102400$, and $N=30720$ and $T=307200$. We can see that with increase in $N$ the shape of the distribution becomes narrower, implying that the amount of spurious cross-correlations decreases. Ideally, the distribution should be a Dirac-delta at zero, since true cross-correlations do not exist. The eigenvalue spectra are less sensitive to the parameters $N$ and $T$, as can be seen in Figs.~\ref{fig_Fig1} (e) and (f), which show the corresponding eigenvalue spectra. For all of the above simulations, we find the simulated data agree closely  with the theoretical Mar\u{c}enko-Pastur distributions (red dashed lines) with $\lambda_{max}=1.732$ and $\lambda_{min}=0.468$ (theoretically calculated using Eq.~\ref{eq_lambda}, and $Q=10$).
%%>>>>>>>>>>>>>>>>>>>>>>>>>>>>>>>>>>>>>>>>>>>>>>>>>>>>>>>>>>>>>>>>>>>>>>>>>>>>>>>>>>>>>>>.
%%%%%*********************************
%%%%%*********************************
%%%%%*********************************
\begin{figure}[!t]
\centering
\includegraphics[width=0.45\linewidth]{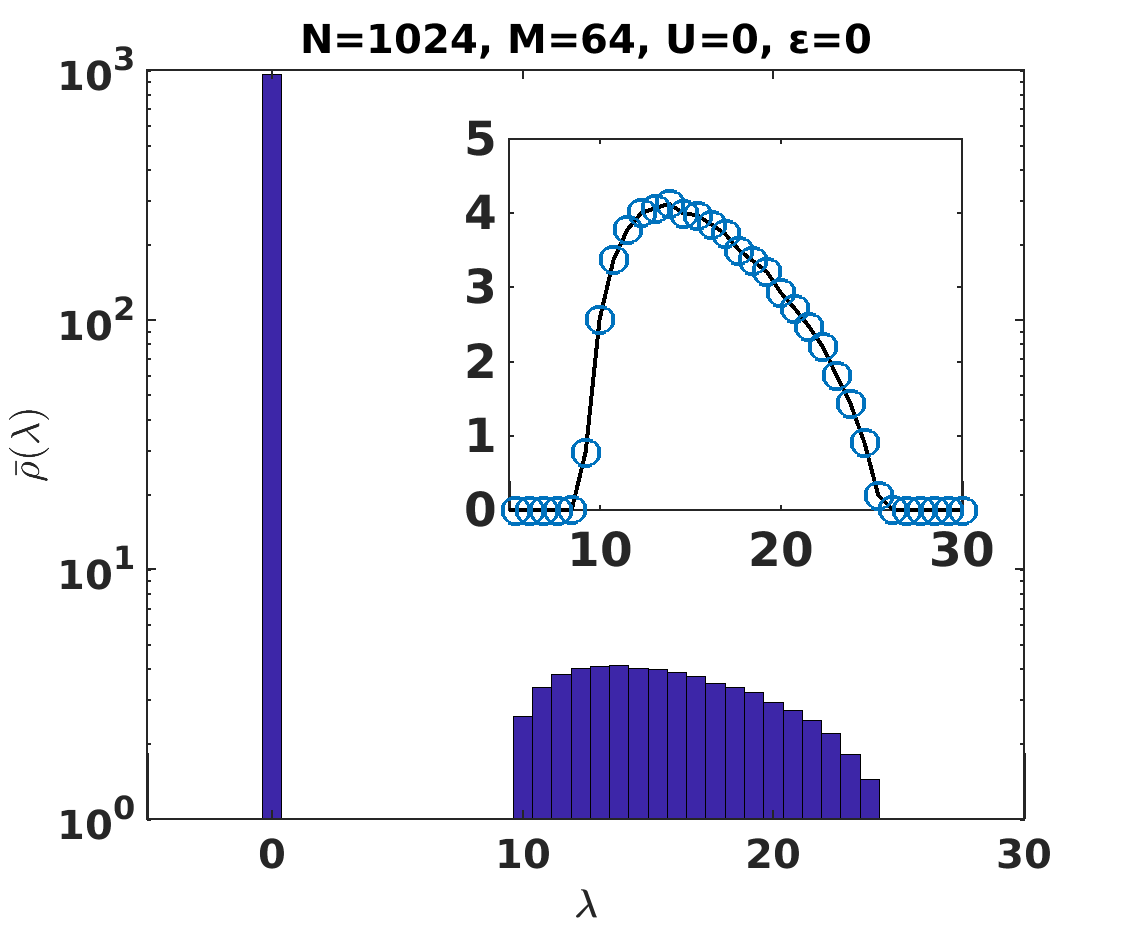}\llap{\parbox[b]{2.1in}{(\textbf{a})\\\rule{0ex}{1.6in}}}\includegraphics[width=0.45\linewidth]{Figures/Fig2b.png}\llap{\parbox[b]{2.1in}{(\textbf{b})\\\rule{0ex}{1.6in}}}
\includegraphics[width=0.45\linewidth]{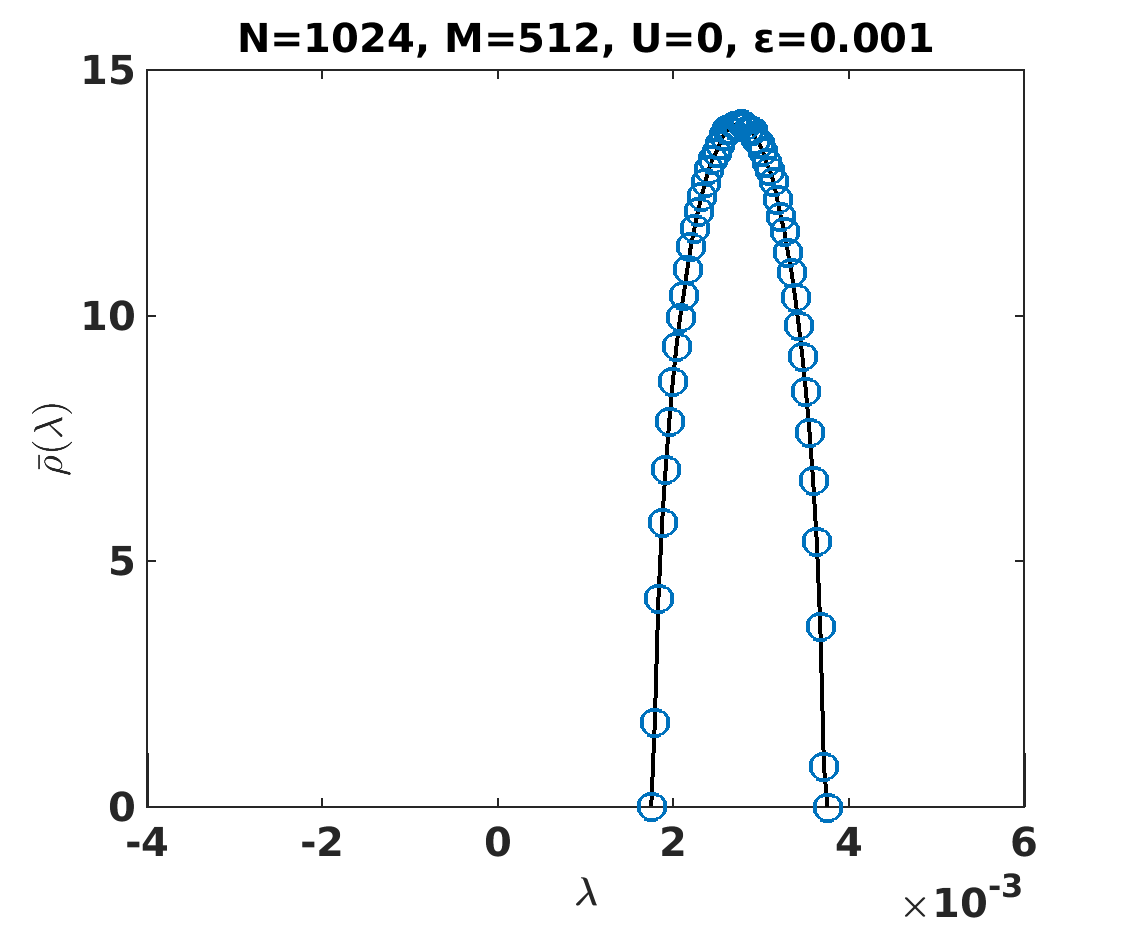}\llap{\parbox[b]{2.1in}{(\textbf{c})\\\rule{0ex}{1.6in}}}\includegraphics[width=0.45\linewidth]{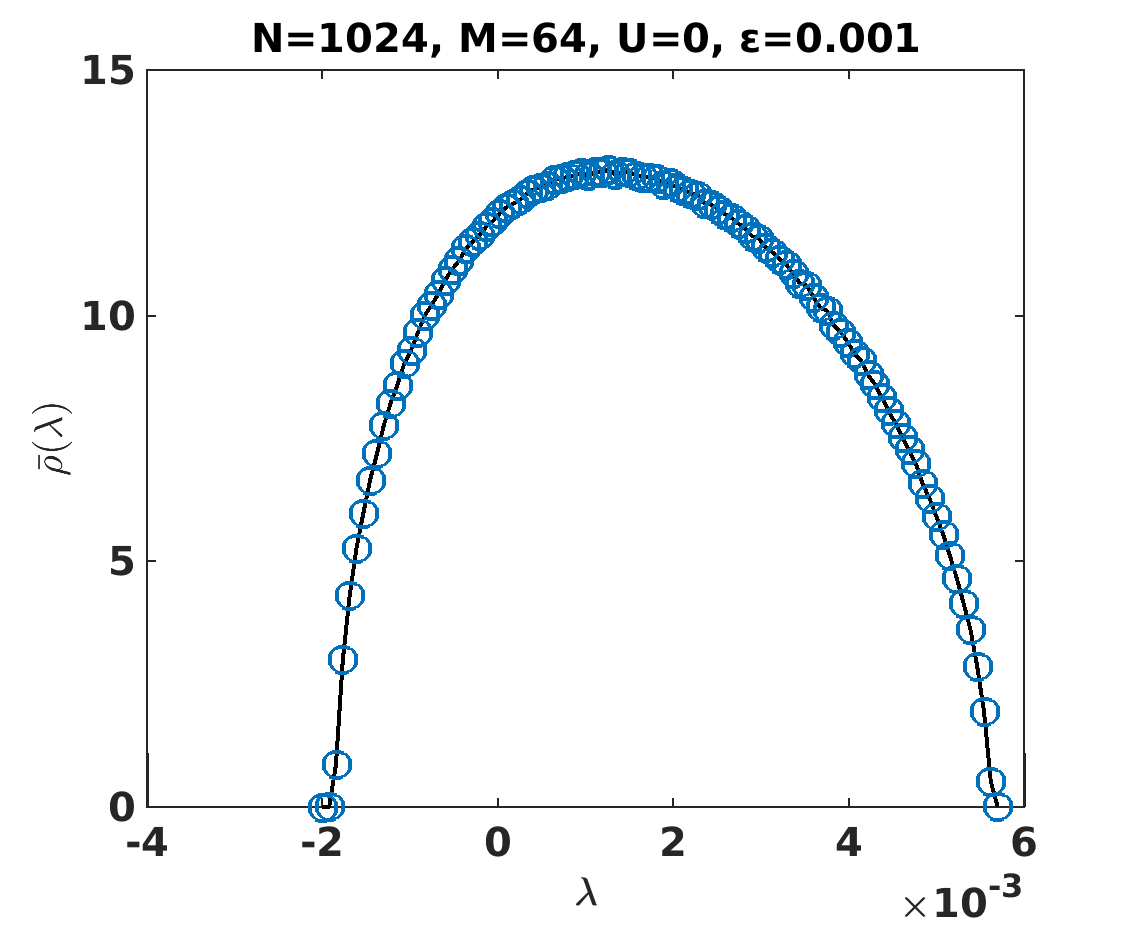}\llap{\parbox[b]{2.1in}{(\textbf{d})\\\rule{0ex}{1.6in}}}
\caption{Semi-log plot of the eigenvalue distribution of Wishart matrix $\boldsymbol W$, using the set of parameters (a) $N=1024$  and $M = 512$; (b) $N = 1024$  and $M = 64$. For short epochs ($N>M$), the eigenvalue spectra have $N-M+1$ zero eigenvalues and the remaining eigenvalues of the spectra show a distributions similar to the Mar\u{c}enko-Pastur distribution. Insets show the zoomed in views of remaining $M-1$ eigenvalues. (c) and (d) show the emerging spectra, generated using the power map technique with $\epsilon = 0.001$, which are deformed semi-circular.  Numerical results for densities of eigenvalues have been generated using the averages over 1000 ensemble members. Note that the emerging spectrum shifts towards left for smaller values of $M$, and also some of its eigenvalues become negative at smaller values of $M$.} 
\label{fig_Fig2}
\end{figure}
%%%%%*********************************

As we have mentioned earlier, the assumption of stationarity fails for a very long return time series, so it is often useful to break one long time series of length $T$ into $n$ shorter epochs, each of size $M$ (such that $T/M=n$). The assumption of stationarity then improves for each of the shorter epochs. However, if there are $N$ return time series, such that $N >> M$, then the corresponding cross-correlation matrices are highly singular with $N-M+1$ zero eigenvalues, which lead to poor eigenvalue statistics. We use the power map technique \cite{Guhr_2003, vinayak_2014} to break the degeneracy of eigenvalues at zero. In this method, a non-linear distortion is given to each element $(W_{ij})$ of the Wishart matrix $\boldsymbol W$ (or later in each correlation coefficient $C_{ij}$ of the empirical cross-correlation matrix $\boldsymbol C$) of short epoch by: 
\begin{equation}\label{wishart_eqn}
W_{ij} \rightarrow (\mathrm{sign} ~~W_{ij}) |W_{ij}|^{1+\epsilon},
\end{equation}
 where $\epsilon$ is a noise-suppression parameter. For very small distortions, e.g., $\epsilon=0.001$ (as used here), we get an ``emerging spectrum'' of eigenvalues, arising from the degenerated eigenvalues at zero which is well separated from the original spectrum. The power mapping method suppresses noise present in the correlation structure of short-time series (see e.g., Refs.~\cite{Chakraborti_2018, Mijail_2018,Ochoa_2018, Pharasi_2018, vinayak_epl} for recent studies and applications). Later in this chapter, we study different aspects of the power mapping method by varying the value of distortion $\epsilon$ from $0$ to $0.8$.

%%%%%*********************************
\begin{figure}[!t]
\centering
\includegraphics[width=0.33\linewidth]{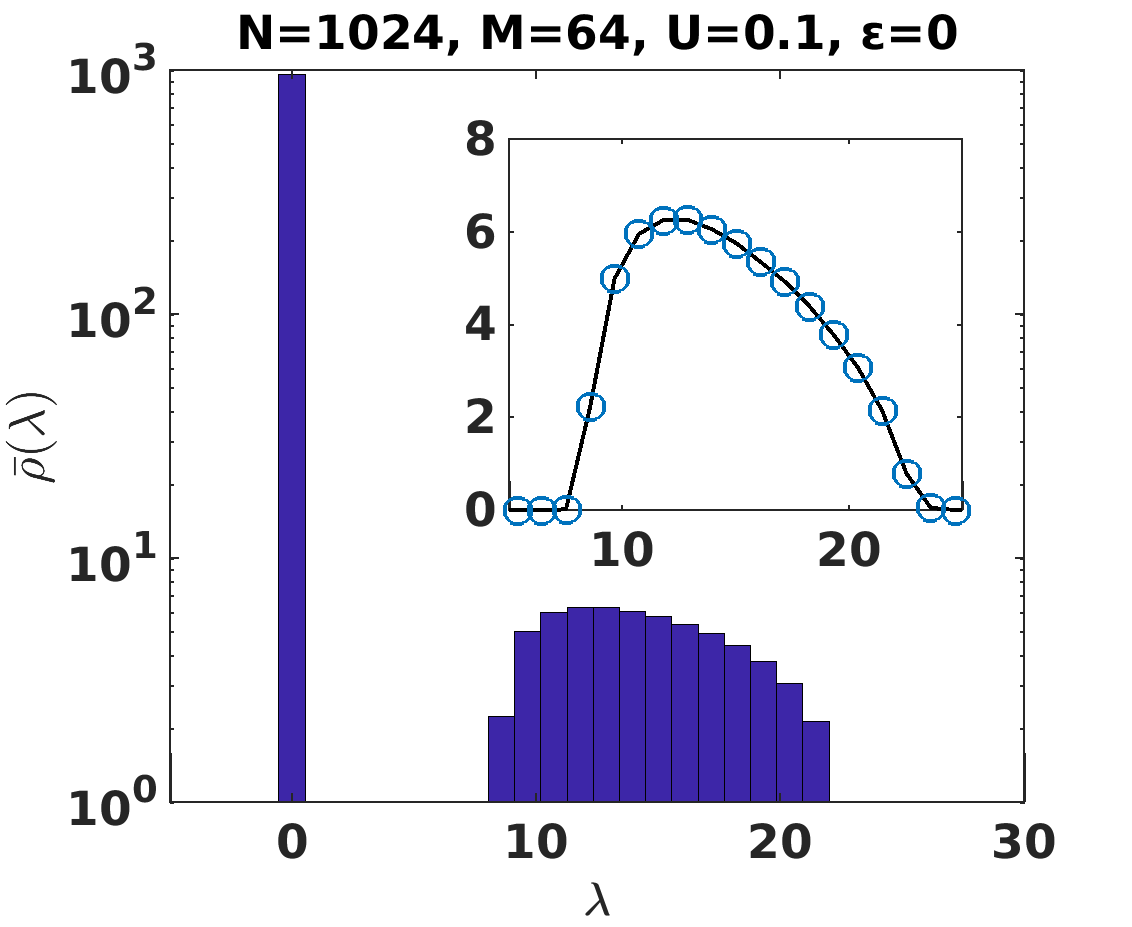}\llap{\parbox[b]{1.4in}{(\textbf{a})\\\rule{0ex}{1.2in}}}\includegraphics[width=0.33\linewidth]{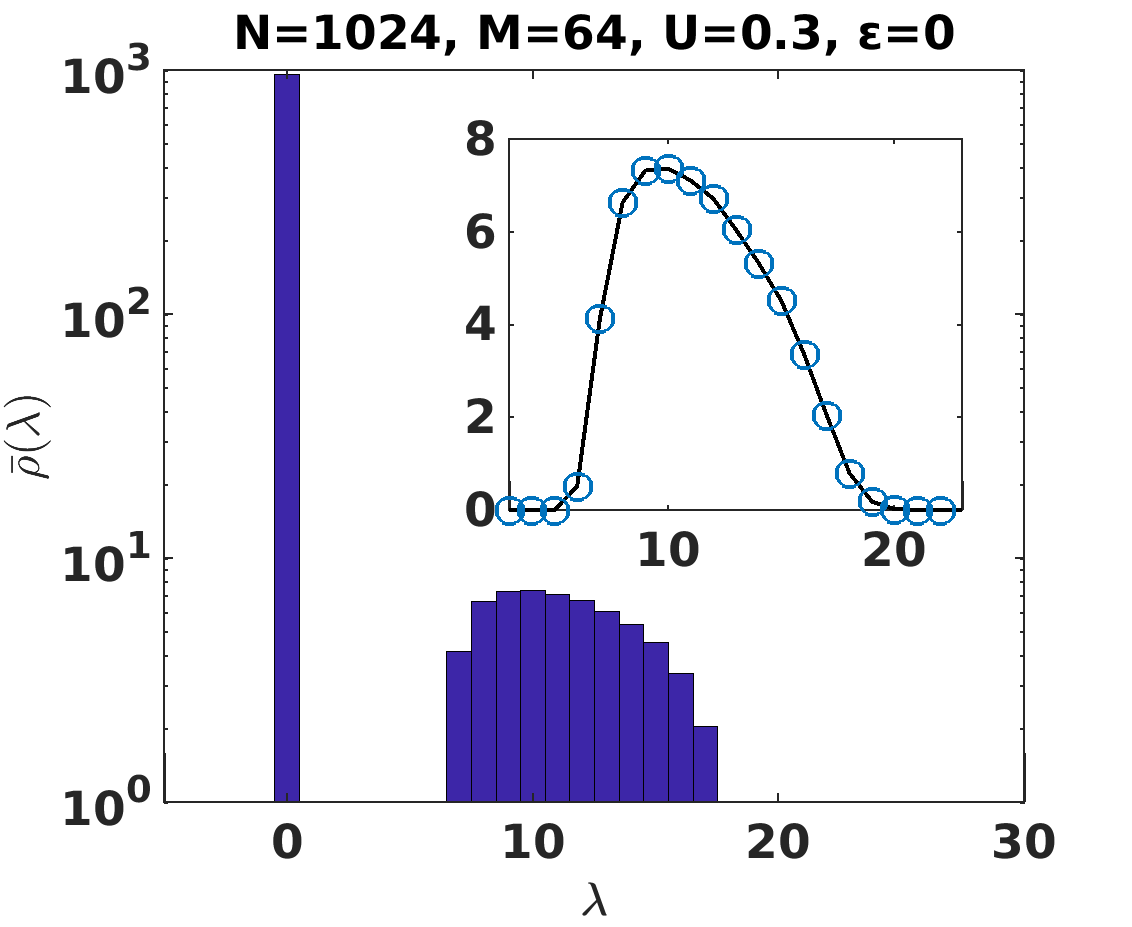}\llap{\parbox[b]{1.4in}{(\textbf{b})\\\rule{0ex}{1.2in}}}\includegraphics[width=0.33\linewidth]{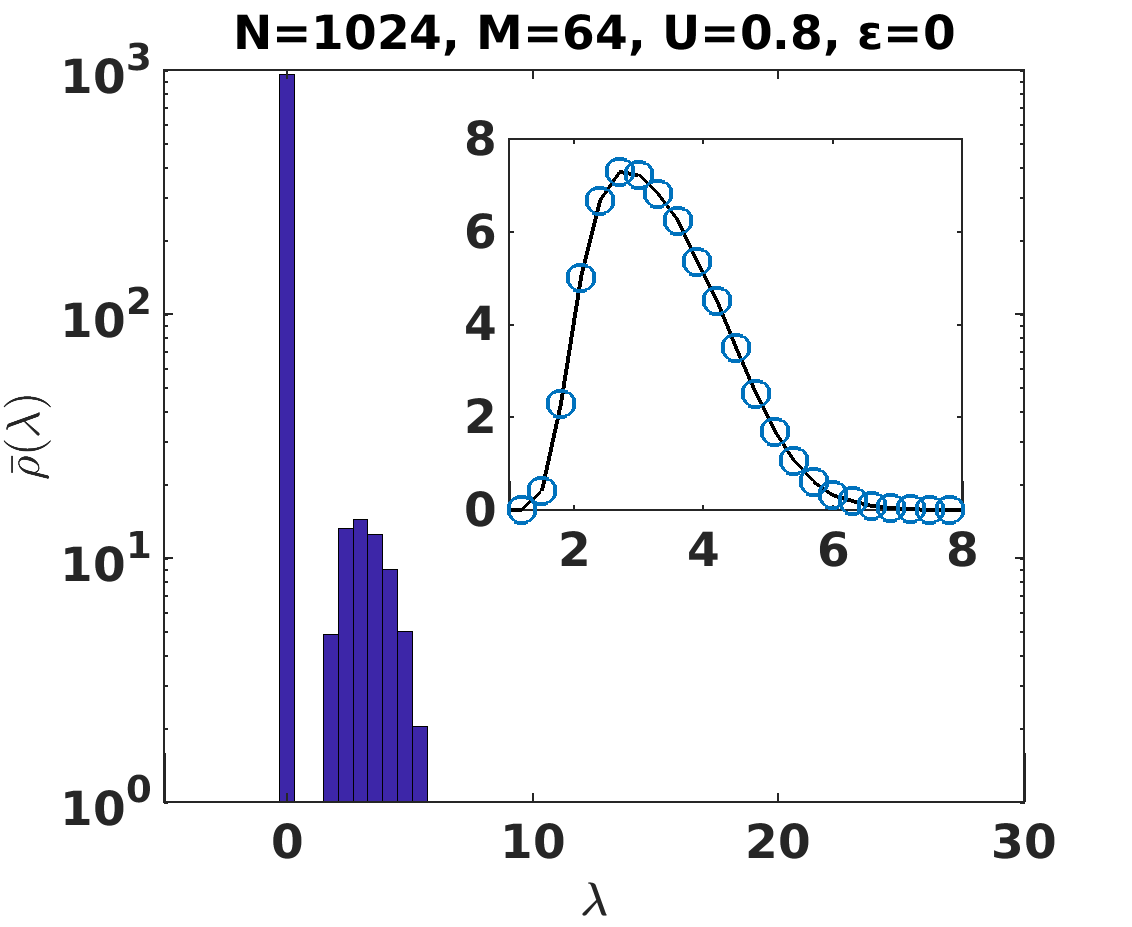}\llap{\parbox[b]{1.4in}{(\textbf{c})\\\rule{0ex}{1.2in}}}\\
\includegraphics[width=0.33\linewidth]{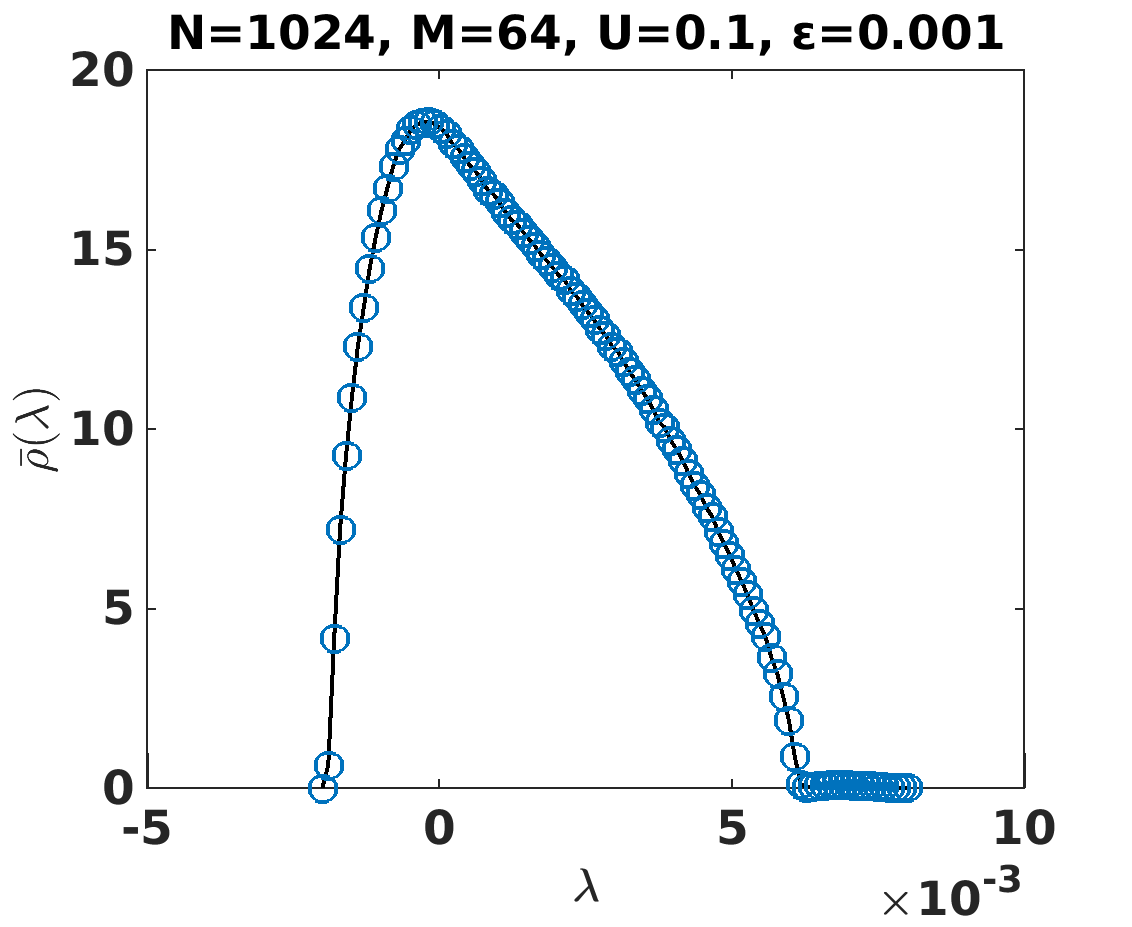}\llap{\parbox[b]{1.4in}{(\textbf{d})\\\rule{0ex}{1.2in}}}\includegraphics[width=0.33\linewidth]{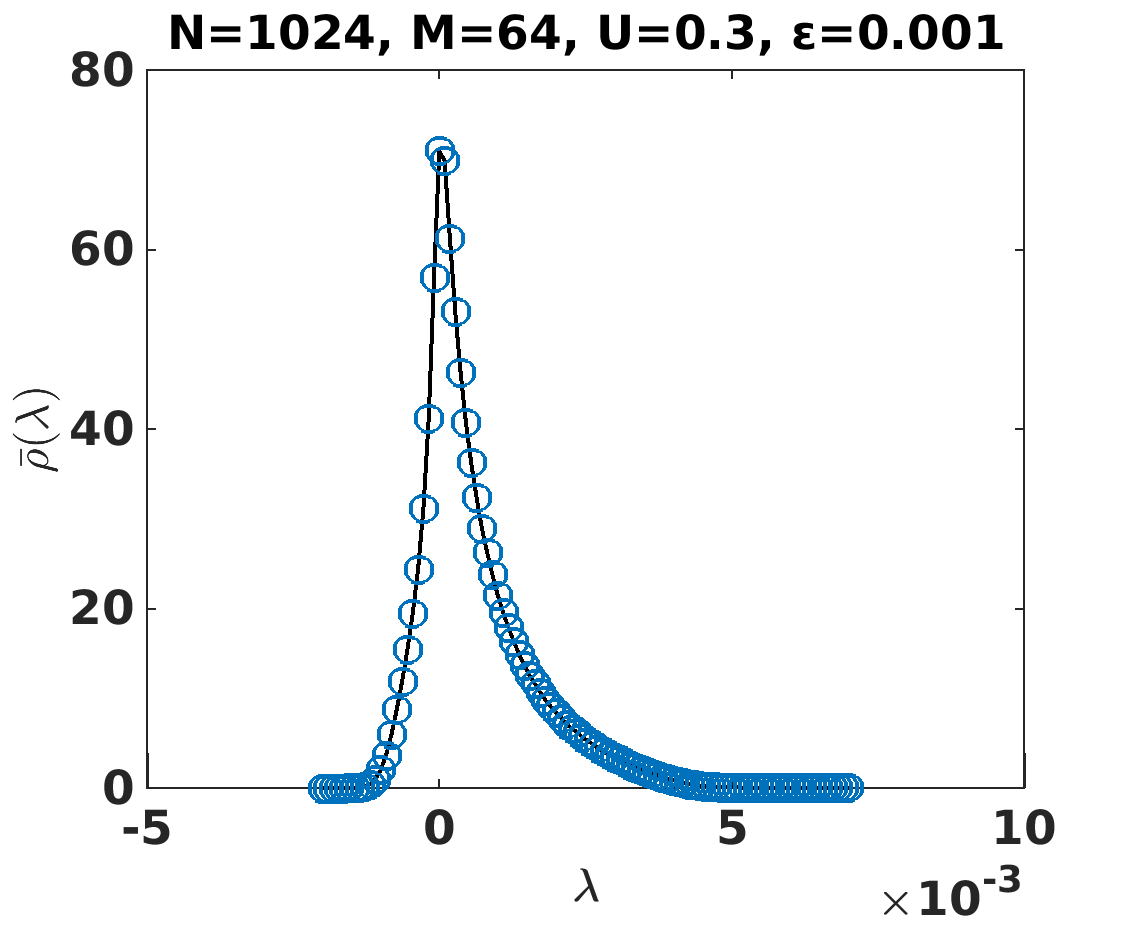}\llap{\parbox[b]{1.4in}{(\textbf{e})\\\rule{0ex}{1.2in}}}\includegraphics[width=0.33\linewidth]{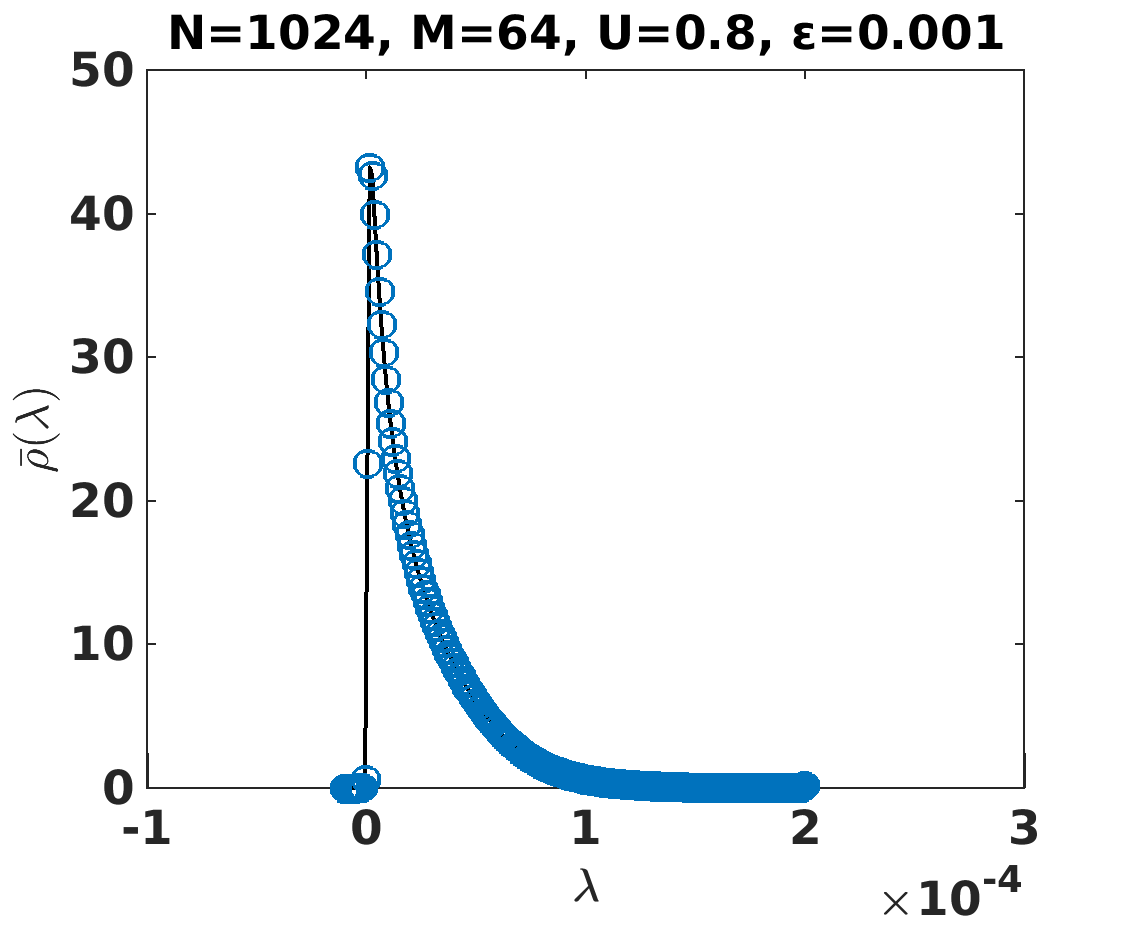}\llap{\parbox[b]{1.4in}{({f})\\\rule{0ex}{1.2in}}}\\
\caption{Eigenvalue spectra of correlated Wishart ensembles with parameters, $N = 1024$ and $M = 64$, shown on semi-log scales with constant correlations: (a) $U=0.1$, (b) $U=0.3$, and (c) $U=0.8$. Insets show the corresponding densities of non-zero eigenvalues, which are closely described by the Mar\u{c}enko-Pastur distributions. (d)-(f) show the densities of the emerging spectra, when non-linear distortions (with $\epsilon = 0.001$) are applied to the same matrices. Note that the shape of the emerging spectrum changes from distorted semi-circle to a Lorentzian-like with the increase of constant correlation strength $U$.} 
\label{fig_CWE1}
\end{figure}
%%%%%%**********************************
In Fig.~\ref{fig_Fig2}, we have studied the effect of non-linear distortion on the behavior of  Wishart ensemble ($U=0$), where $N>>M$. The top row of Fig.~\ref{fig_Fig2} shows semi-log plots of the ensembles with parameters: (a) $N=1024$ and $M=512$, and (b) $N=1024$ and $M=64$. Then small non-linear distortions with $\epsilon=0.001$ are given to the ensembles to display the emerging spectra,  shown in Figs.~\ref{fig_Fig2} (c) and (d). Interestingly, the shape of the emerging spectrum changes from a semi-circle to a strongly distorted one, as $M$ becomes shorter. Also, note that emerging spectrum shifts towards the left side as $M$ becomes shorter. For smaller values of $M$, some of the eigenvalues of emerging spectrum become negative. The number of negative eigenvalues depend on the size of the epoch $M$, the distortion parameter $\epsilon$ and the mean correlation in the case of a correlated Wishart ensemble \cite{Ochoa_2018}.

%%%*****************************************
\begin{figure}[!t]
\centering
\includegraphics[width=0.33\linewidth]{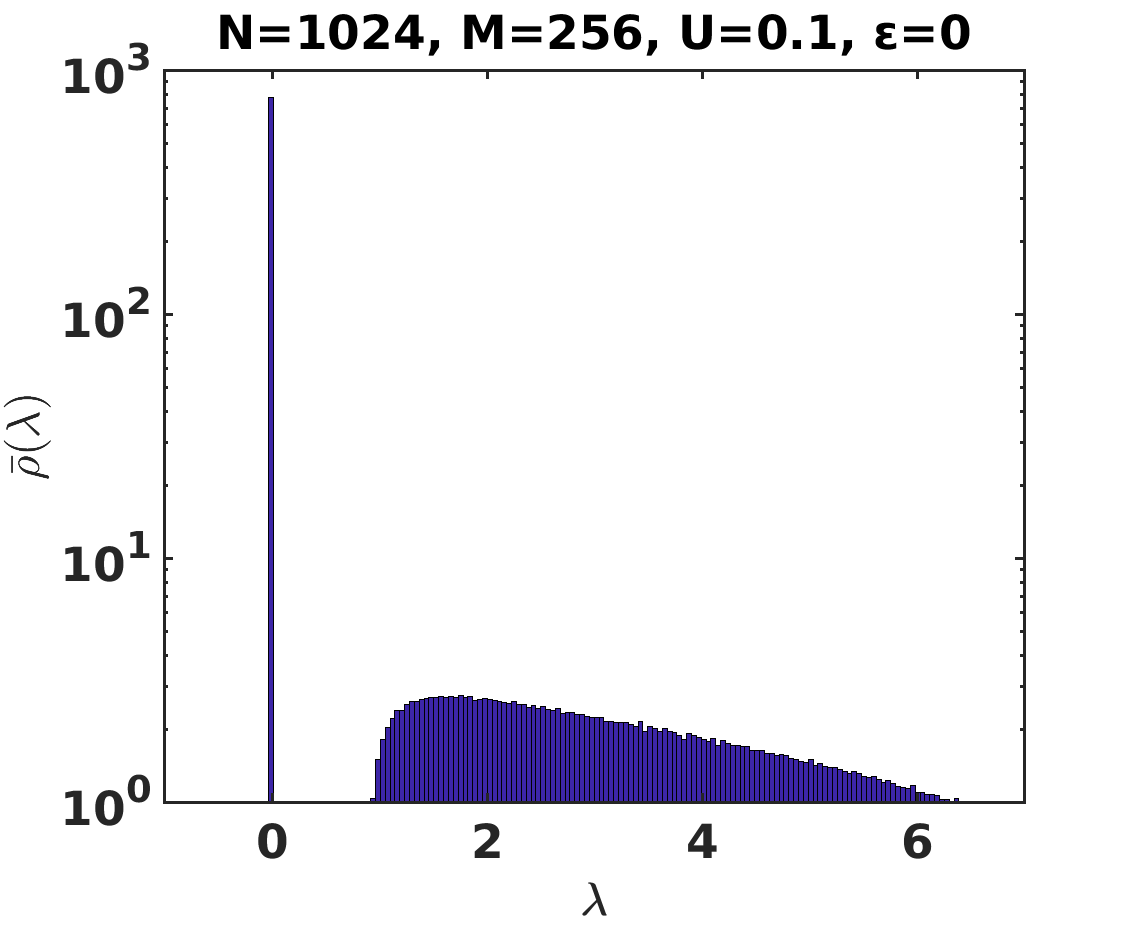}\llap{\parbox[b]{1.4in}{(\textbf{a})\\\rule{0ex}{1.2in}}}\includegraphics[width=0.33\linewidth]{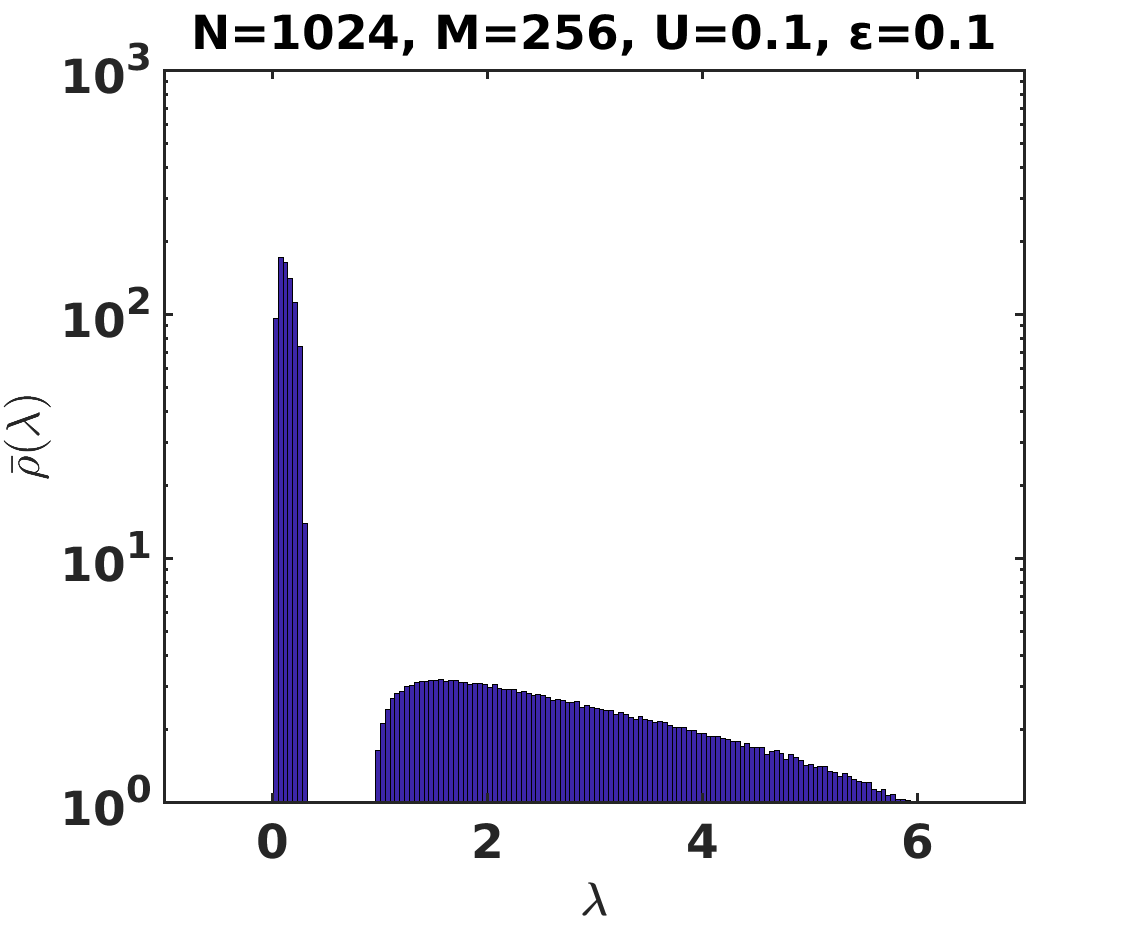}\llap{\parbox[b]{1.4in}{(\textbf{b})\\\rule{0ex}{1.2in}}}\includegraphics[width=0.33\linewidth]{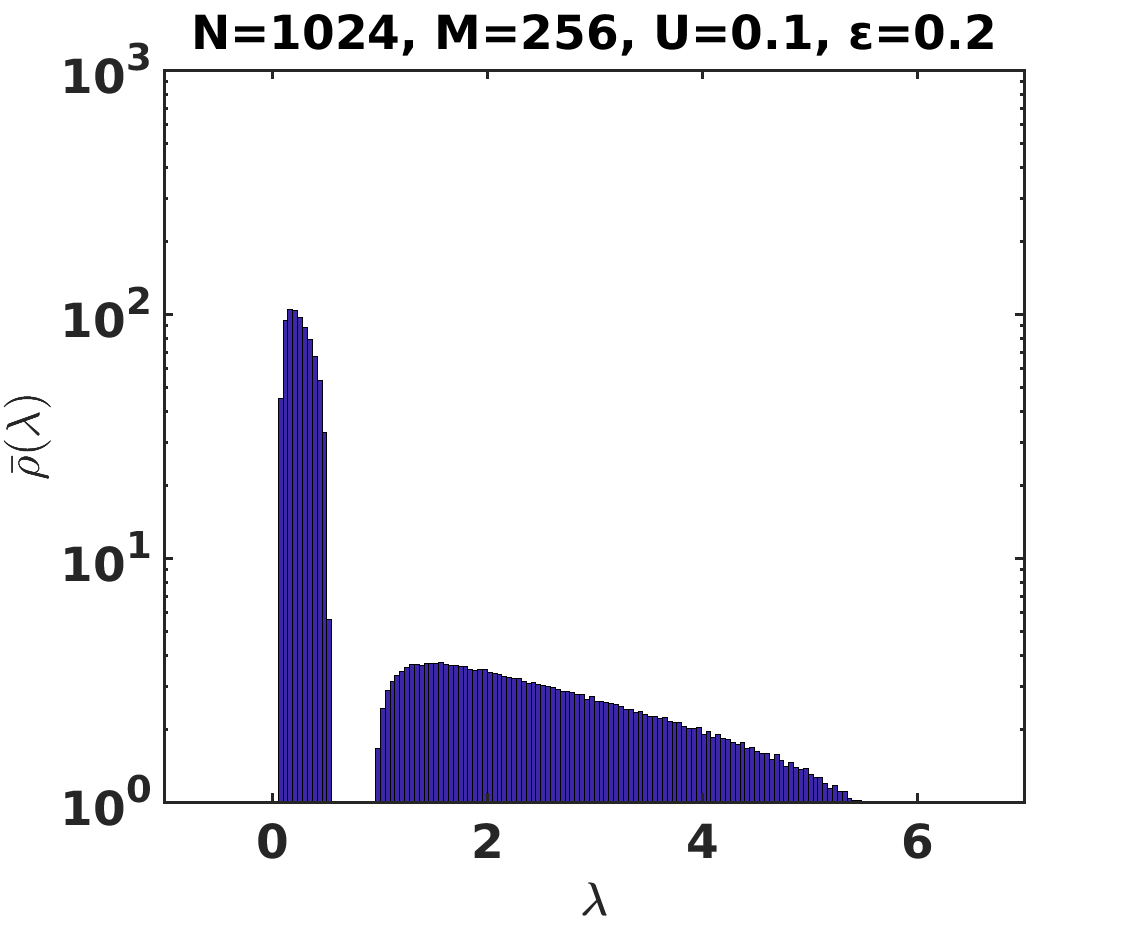}\llap{\parbox[b]{1.4in}{(\textbf{c})\\\rule{0ex}{1.2in}}}\\
\includegraphics[width=0.33\linewidth]{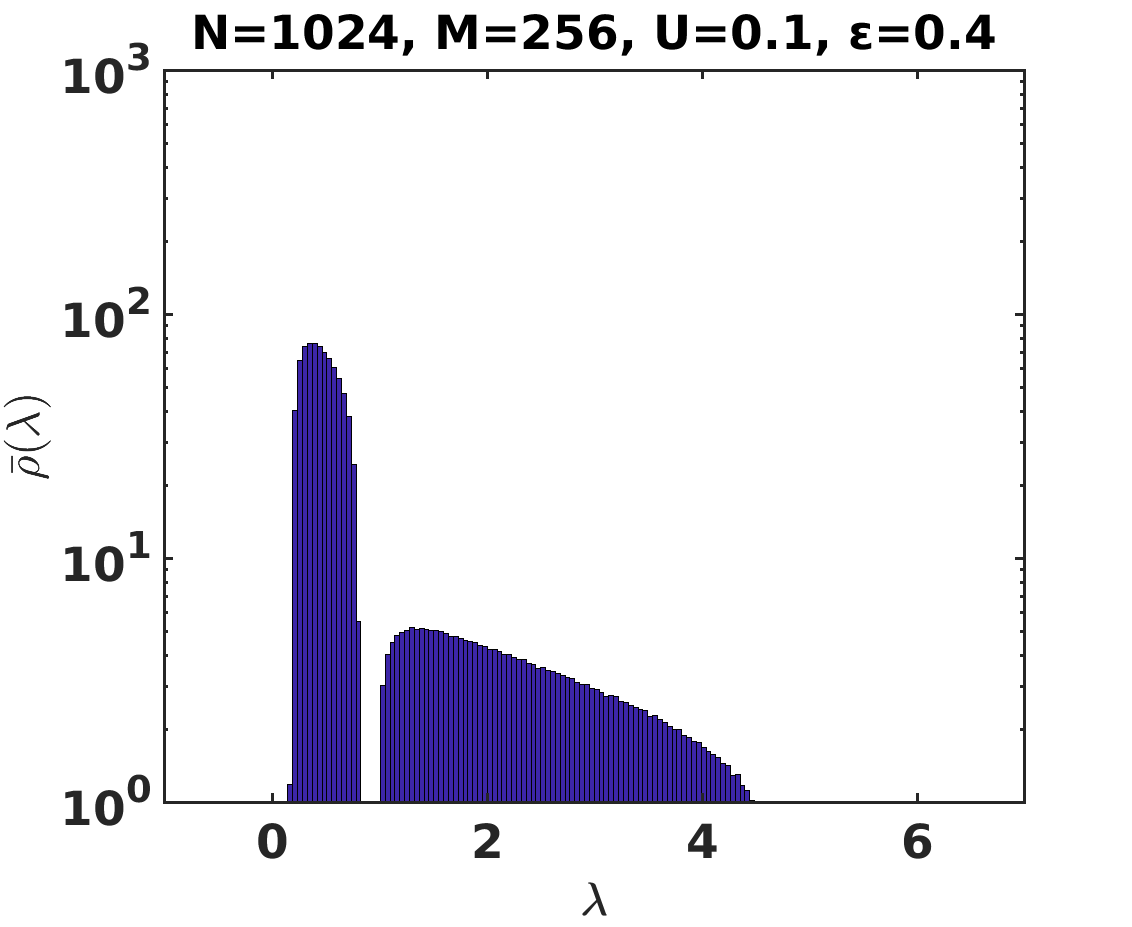}\llap{\parbox[b]{1.4in}{(\textbf{d})\\\rule{0ex}{1.2in}}}\includegraphics[width=0.33\linewidth]{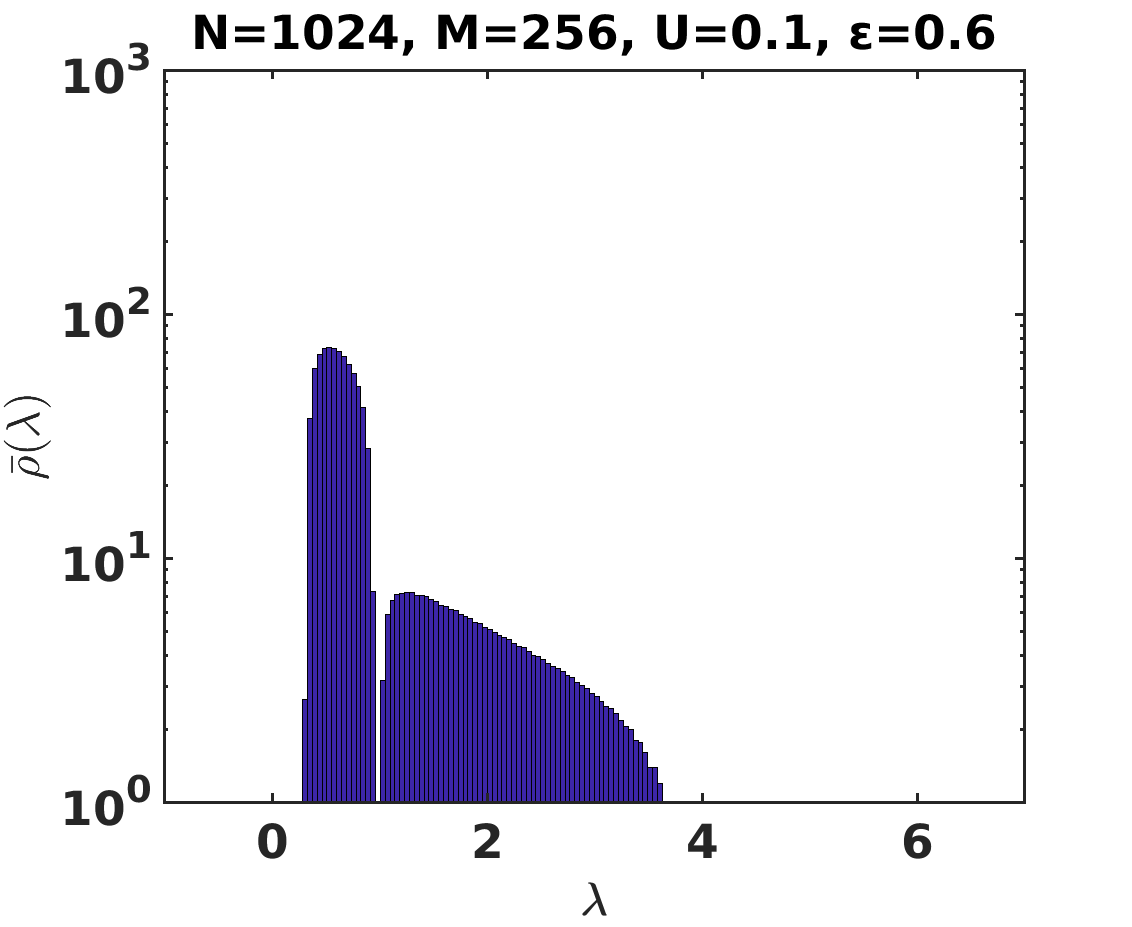}\llap{\parbox[b]{1.4in}{(\textbf{e})\\\rule{0ex}{1.2in}}}\includegraphics[width=0.33\linewidth]{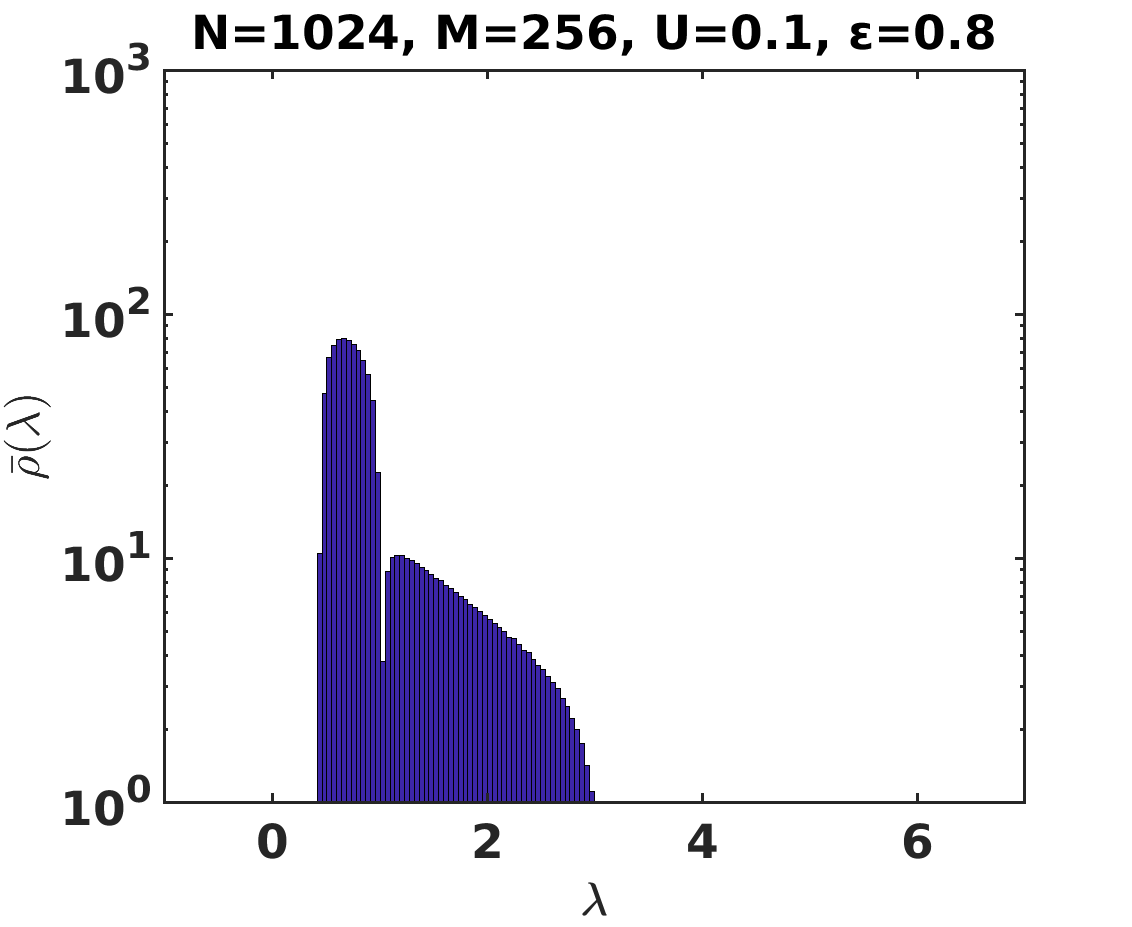}\llap{\parbox[b]{1.4in}{(\textbf{f})\\\rule{0ex}{1.2in}}}\\
\caption{ Semi-log plots of the eigenvalue spectra for the correlated Wishart ensemble $\boldsymbol W$ with parameters $N = 1024$ and $M = 256$  at a constant correlation with $U=0.1$, and distortion parameters of: (a) $\epsilon=0$, (b) $\epsilon=0.1$, (c) $\epsilon=0.2$, (d) $\epsilon=0.4$, (e) $\epsilon=0.6$, and (f) $\epsilon=0.8$. For $\epsilon=0.1$, the emerging spectrum is well separated from non-zero eigenvalues but with the increase of the distortion parameter $\epsilon$ the emerging spectrum starts moving towards the remaining non-zero eigenvalues spectra, and eventually merges with it at higher values, e.g., $\epsilon=0.8$.} 
\label{fig_CWE2}
\end{figure}
%%%*****************************************
Fig.~\ref{fig_CWE1} shows the effect of a constant correlation with strength $U$ on the eigenvalue spectra and the emerging spectra of correlated Wishart ensembles with parameters $N = 1024$ and $M = 64$. Figs.~\ref{fig_CWE1} (a)-(c) show the eigenvalue distributions, on the semi-log scales, for the correlated Wishart ensembles with correlations $U=0.1$,  $U=0.3$, and $U=0.8$, respectively. Insets show the densities of non-zero eigenvalues, which are closely described by the Mar\u{c}enko-Pastur distributions in all cases. In the bottom row, Figs.~\ref{fig_CWE1} (d)-(f) show the densities of the corresponding emerging spectra arising from non-linear distortion of the degenerate eigenvalues at zero. The shapes of the emerging spectra change from distorted semi-circle to Lorentzian-like, as the constant correlation values increase for the correlated Wishart ensembles.

Next, we present the effect of the distortion (or noise-suppression) parameter $\epsilon$ on the eigenvalue spectra in Fig.~\ref{fig_CWE2}. Fig.~\ref{fig_CWE2} (a)-(f) show the distributions of eigenvalues for the correlated Wishart ensembles with parameters $ N=1024$ and $M=64$, and varying distortion parameter values: $ \epsilon=0.0, 0.1,0.2,0.4,0.6$ and $0.8$, keeping a constant correlation ($U=0.1$) among all off-diagonal elements in $\boldsymbol \zeta$. The densities of non-zero eigenvalues are closely described by the Mar\u{c}enko-Pastur distributions, but the emerging spectra move towards the main spectra as the value of $\epsilon$ increases. The emerging spectra is absent at $\epsilon =0$, while it merges with the main spectrum at high values of distortion parameter, e.g., $\epsilon =0.8$.

%%>>>>>>>>>>>>>>>>>>>>>>>>>>>>>>>>>>>>>>>>>>>>>>>>>>>>>>>>>>>>>>>>>>>>>>>>>>>>>>>>>>>>>.
\subsubsection{Eigenvalue decomposition of the empirical cross-correlation matrix}
\begin{figure}[!b]
\centering
\includegraphics[width=2.4cm, height=2.2cm]{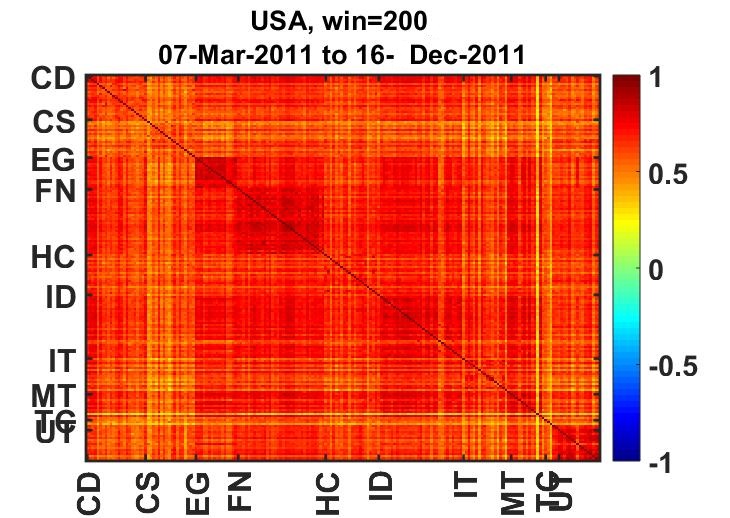}
\llap{\parbox[b]{1.1in}{(\textbf{a})\\\rule{0ex}{0.7in}}}
\includegraphics[width=2.4cm, height=2.2cm]{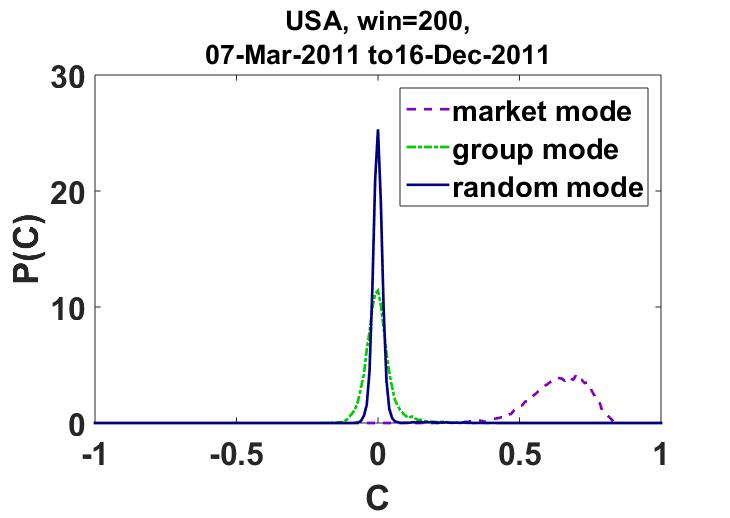}
\llap{\parbox[b]{1.1in}{(\textbf{b})\\\rule{0ex}{0.7in}}}
\includegraphics[width=2.4cm, height=2.2cm]{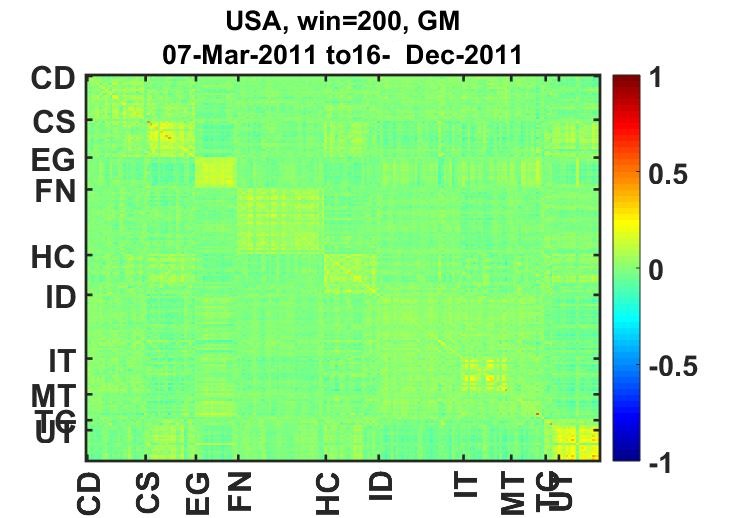}
\llap{\parbox[b]{1.1in}{(\textbf{c})\\\rule{0ex}{0.7in}}}
\includegraphics[width=2.4cm, height=2.2cm]{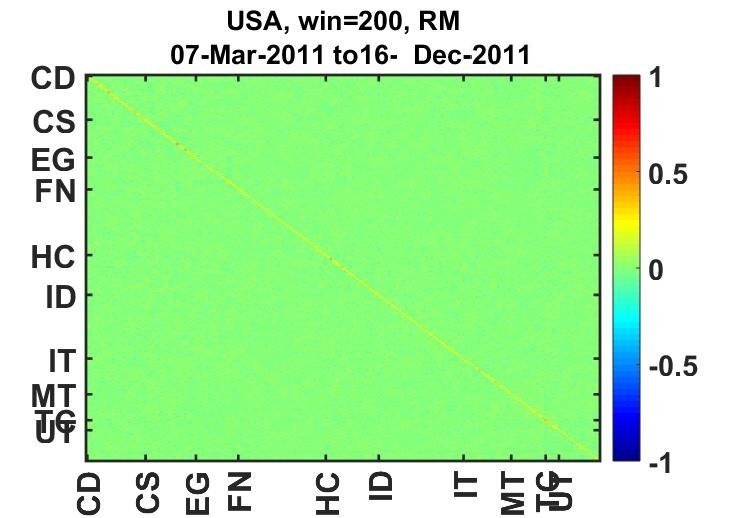}
\llap{\parbox[b]{1.1in}{(\textbf{d})\\\rule{0ex}{0.7in}}}
\includegraphics[width=2.4cm, height=2.2cm]{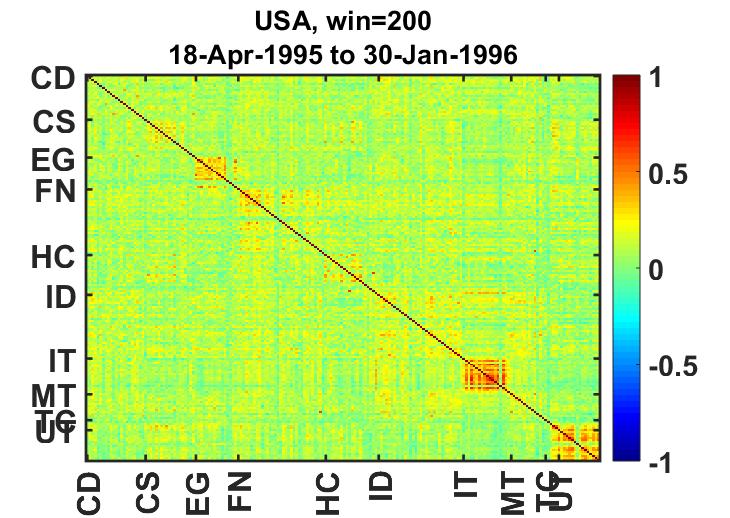}
\llap{\parbox[b]{1.1in}{(\textbf{e})\\\rule{0ex}{0.7in}}}
\includegraphics[width=2.4cm, height=2.2cm]{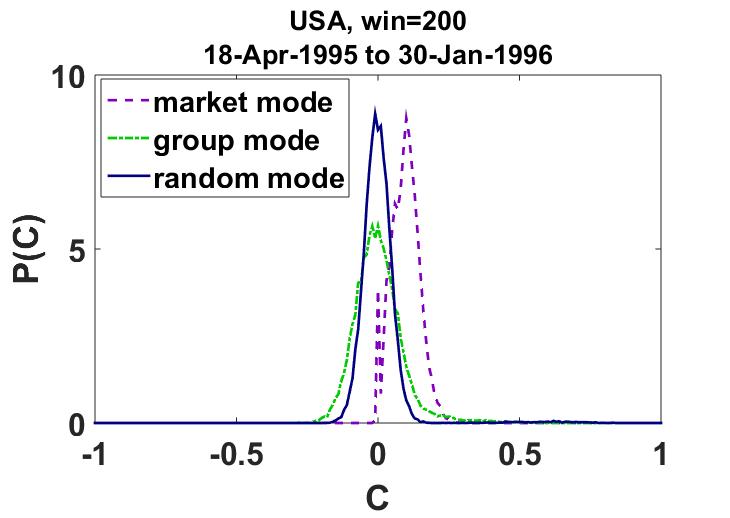}
\llap{\parbox[b]{1.1in}{(\textbf{f})\\\rule{0ex}{0.7in}}}
\includegraphics[width=2.4cm, height=2.2cm]{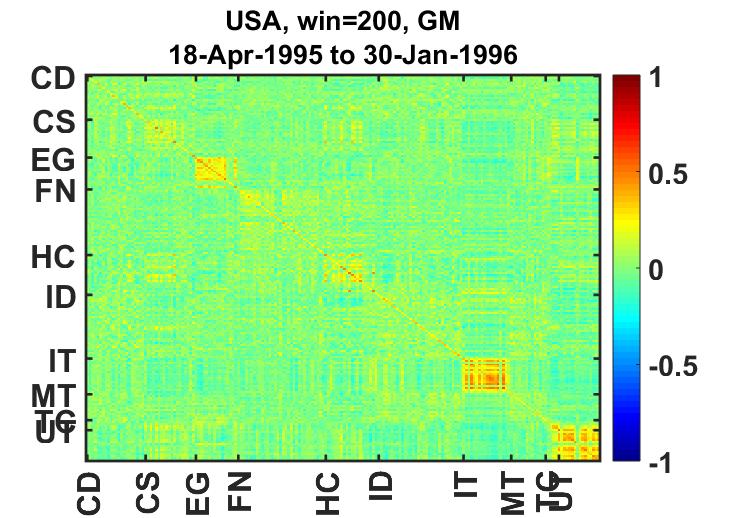}
\llap{\parbox[b]{1.1in}{(\textbf{g})\\\rule{0ex}{0.7in}}}
\includegraphics[width=2.4cm, height=2.2cm]{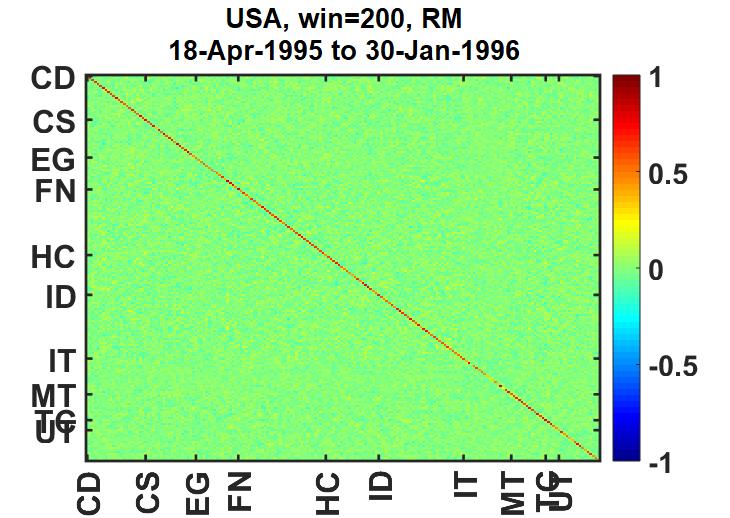}
\llap{\parbox[b]{1.1in}{(\textbf{h})\\\rule{0ex}{0.7in}}}

\caption{(a) and (e) show the cross-correlation matrices of $194$ stocks of S\&P 500  for $M=200$ days during: (a) 07/03/2011 to 16/12/2011; (b) 18/04/1995 to 30/01/1996. The stocks are arranged according to their industrial groups (abbreviations are given in Table~\ref{Table:sectoral_index_sp500}). 
The blocks along the diagonal show the correlations within the same industrial groups; the color-bar shows the amount of correlation among the stocks. (a) shows the correlation matrix with high mean correlation and (e) with low mean correlation. (b) and (f) show the eigenvalue decomposition of the correlation matrix into the market mode, group modes and random modes. The market mode captures the mean market correlation, which is the dominant eigenvalue of the matrix. The group modes give the sectoral behavior of the market characterized by the subsequent 15 eigenvalues for a correlation matrix (a) and the next 62 eigenvalues for a correlation matrix (e) of the market. The rest of the eigenvalues show random behavior. (c) and (g) are the correlation matrix after removing the market mode  and random modes from the correlation matrix; thus the matrix is composed of group modes only. We can visualize the block structure which shows the correlation among sectors. (d) and (h) show the correlation matrix after removing the market mode and group modes from the correlation matrix; so the matrix is composed of random modes only.}
\label{fig_USA_corr_decomp}
\end{figure}

We also analyze $N=194$ adjusted daily closure price time series of the stocks of S\&P 500 (USA) index from the Yahoo finance database \cite{Yahoo_finance}. As discussed in the methodology subsection, we construct the empirical cross-correlation matrix $\boldsymbol C(\tau)$ for an epoch of $M=200$ trading days, ending on trading day $\tau$. In Fig.~\ref{fig_USA_corr_decomp} (a) and (e), we choose two correlation matrices for the time series from 07/03/2011 to 16/12/2011 (high mean correlation) and  18/04/1995 to 30/01/1996 (low mean correlation), respectively.  The color-bar shows the amount of correlation among the stocks. The stocks are arranged according to their industrial groups (abbreviations are given in Table~\ref{Table:sectoral_index_sp500}). The blocks along the diagonal show the correlations within the same industrial group.  Fig.~\ref{fig_USA_corr_decomp} (b) and (f) show the eigenvalue decomposition of the correlation matrices into the respective market mode, the group modes and the random modes. From such a segregation/decomposition, it is also possible to reconstruct the contributions of different modes to the aggregate correlation matrix as we show below.
%%%*****************************************
%%%%%*********************************

The largest eigenvalue of the correlation matrix, corresponds to a market mode reflects the aggregate dynamics of the market common across all stocks, and strongly correlated to the mean market correlation. The group modes capture the sectoral behavior of the market, which are 15 eigenvalues subsequent to the largest eigenvalue of the correlation matrix of Fig.~\ref{fig_USA_corr_decomp} (c), and the  62 subsequent eigenvalues for correlation matrix of Fig.~\ref{fig_USA_corr_decomp}(g). Remaining eigenvalues capture the random modes behavior of the market (see Fig.~\ref{fig_USA_corr_decomp} (d) and (h)). By using the eigenvalue decomposition, we can thus filter   the true correlations (coming from the signal) and the spurious correlations (coming from the random noise). For this, we first decompose the aggregate correlation matrix as
\begin{equation}
\boldsymbol C = \sum_{i=1}^{N} \lambda_{i}a_{i}a_{i}' ,
\end{equation}
where $\lambda_{i}$ and $a_{i}$ are the eigenvalues and eigenvectors, respectively, of the correlation matrix $\boldsymbol C$. An easy way of handling the reconstruction of the correlation matrix is to sort the eigenvalues in descending order, and then rearranging the eigenvectors in corresponding ranks. This allows one to decompose the matrix into three separate components, viz., market, group and random
\begin{eqnarray}
C &=& C^{M} + C^{G} + C^{R},\\
   &=& \lambda_{1}a_{1}a_{1}' + \sum_{i=2}^{N_{G}} \lambda_{i}a_{i}a_{i}' + \sum_{i=N_{G}+1}^{N} \lambda_{i}a_{i}a_{i}' ,
\end{eqnarray}
where $N_{G}$ is taken to be 15 for the high mean correlated matrix (Fig.~\ref{fig_USA_corr_decomp}(a)) and 62 for the low mean correlation (Fig.~\ref{fig_USA_corr_decomp}(e)), i.e., corresponding to the 15 (or 62) eigenvalues after the largest one, for  two chosen correlation matrices. It is worth noting that the result is not extremely sensitive to the exact value of $N_G$. As mentioned above, the eigenvectors from 2 to $N_{G}$ describe the sectoral dynamics.
%%%%%*****************************************
\begin{figure}[!b]
\centering
\includegraphics[width=0.32\linewidth]{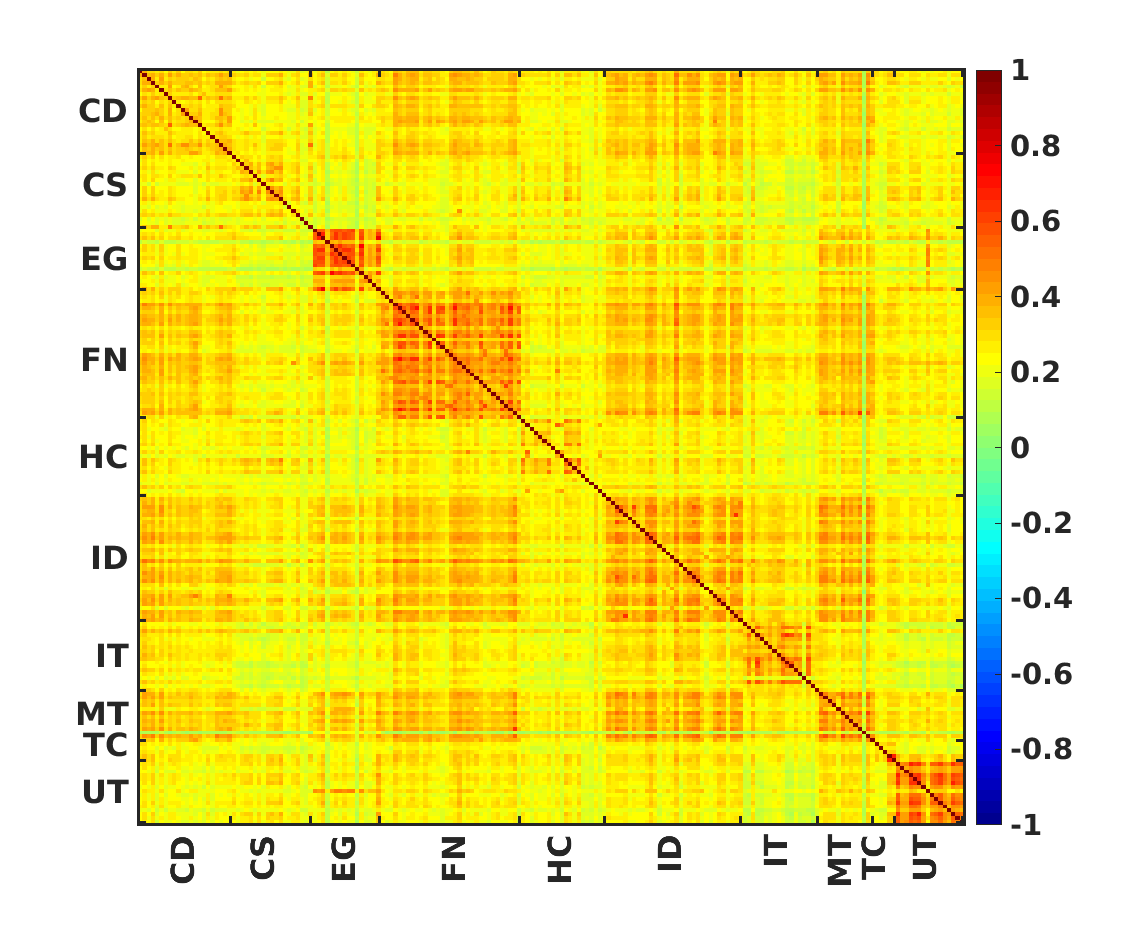}
\llap{\parbox[b]{1.6in}{(\textbf{a})\\\rule{0ex}{1.1in}}}
\includegraphics[width=0.33\linewidth]{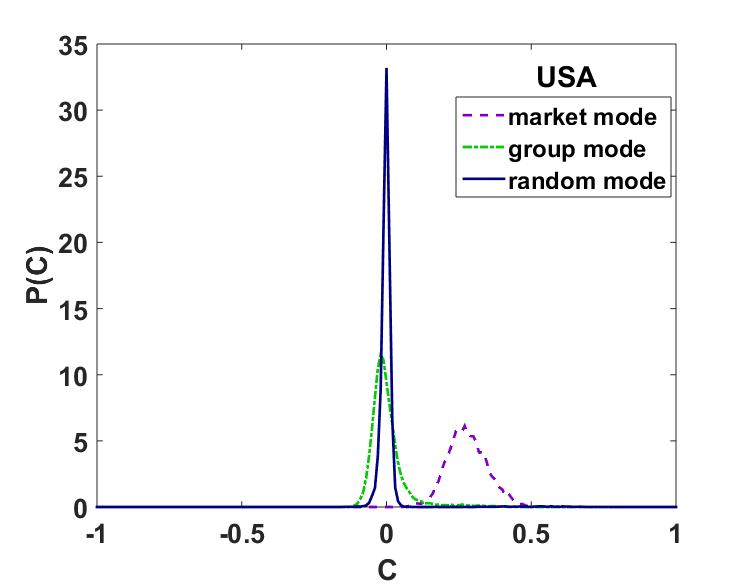}
\llap{\parbox[b]{1.6in}{(\textbf{b})\\\rule{0ex}{1.1in}}}
\includegraphics[width=0.32\linewidth]{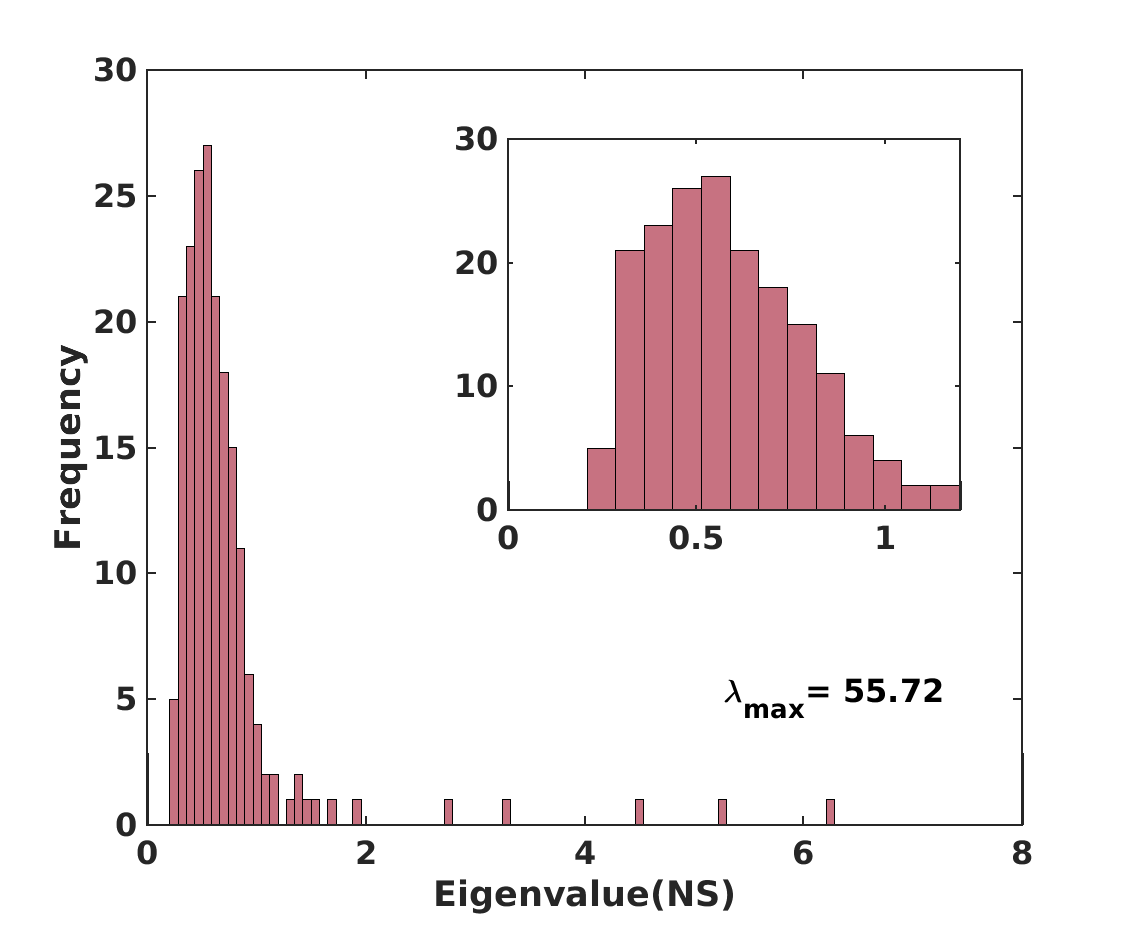}
\llap{\parbox[b]{1.6in}{(\textbf{c})\\\rule{0ex}{1.1in}}}
\caption{(a) Average cross-correlation matrix of $194$ stocks of S\&P 500 in $32$-years period from 1985 to 2016. The stocks are arranged according to their industrial groups (abbreviations are given in Table~\ref{Table:sectoral_index_sp500}). The diagonal blocks show the correlations within the same industrial groups and off diagonal elements show correlations with other industrial groups.
(b) Eigenvalue decomposition of the average correlation matrix into market mode, group modes and random modes. The market mode captures the mean market correlation. The group modes give the sectoral behavior of the market. The random modes of the correlation matrix yield the Mar\u{c}enko-Pastur distribution.
(c) Eigenvalue spectrum of the correlation matrix, evaluated for the \textit{long} return time series for the entire period of $32$-years, with the maximum eigenvalue of the normal spectrum $\lambda_{max}= 55.72$. The largest eigenvalue is well separated from the `bulk'. Inset shows the random part of the spectrum, with the smallest eigenvalue of the normal spectrum $\lambda_{min}= 0.22$.}
\label{fig_USA_avg_corr}
\end{figure}
%%%%%*********************************

%%%%%*********************************
Fig.~\ref{fig_USA_corr_decomp}(c) and (g) show the correlation matrices after removing the market mode  and random modes from the respective correlation matrices; so the matrices show group modes only. We can see the block structures, which exhibit the correlations among the sectors. Figs.~\ref{fig_USA_corr_decomp} (d) and (h) show the correlation matrices after removing the market mode and group modes; so the matrices display the random modes only.

An important observation is that the market mode shifts towards the right with the increment of the mean correlation. The group modes almost coincide with the random modes but with higher variance. Thus, the sectoral dynamics are almost absent whereas the market mode is very strong (similar to what was observed in Ref. \cite{Sharma_2017b}). 

Fig.~\ref{fig_USA_avg_corr} (a) shows the average cross-correlation matrix of $N=194$ stocks of S\&P 500 for the entire duration 1985-2016 ($T=8068$ trading days). We decomposed the average cross-correlation matrix into  the market mode, group modes and random modes. As usual, the market mode captures the mean market correlation corresponding to the maximum eigenvalue, which is  separate from rest of the eigenvalues (see Ref.~\cite{Manan_2018} for the comparison of the behavior of maximum eigenvalues in correlated Wishart ensembles). The group modes, which tell about the sectoral behavior of the market, largely coincide with the random modes and correspond to the random behavior of the stocks. The resulting eigenvalue distribution (shown in Fig.~\ref{fig_USA_avg_corr} (c)) thus has part that is a Mar\u{c}enko-Pastur distribution \cite{Marcenko_1967} (see Fig.~\ref{fig_USA_avg_corr} (c) and its inset) and some deviations. As $ N<<T$ so we do not get any zero eigenvalues. The maximum eigenvalue ($\lambda_{max}= 55.72$) of the spectra dominates the whole market. The next 19 eigenvalues correspond to the group modes, and the rest behave as random modes. The smallest eigenvalue of the spectrum $\lambda_{min}= 0.22$.
%%%%%*********************************
\begin{figure}[!b]
\centering
\includegraphics[width=0.59\linewidth]{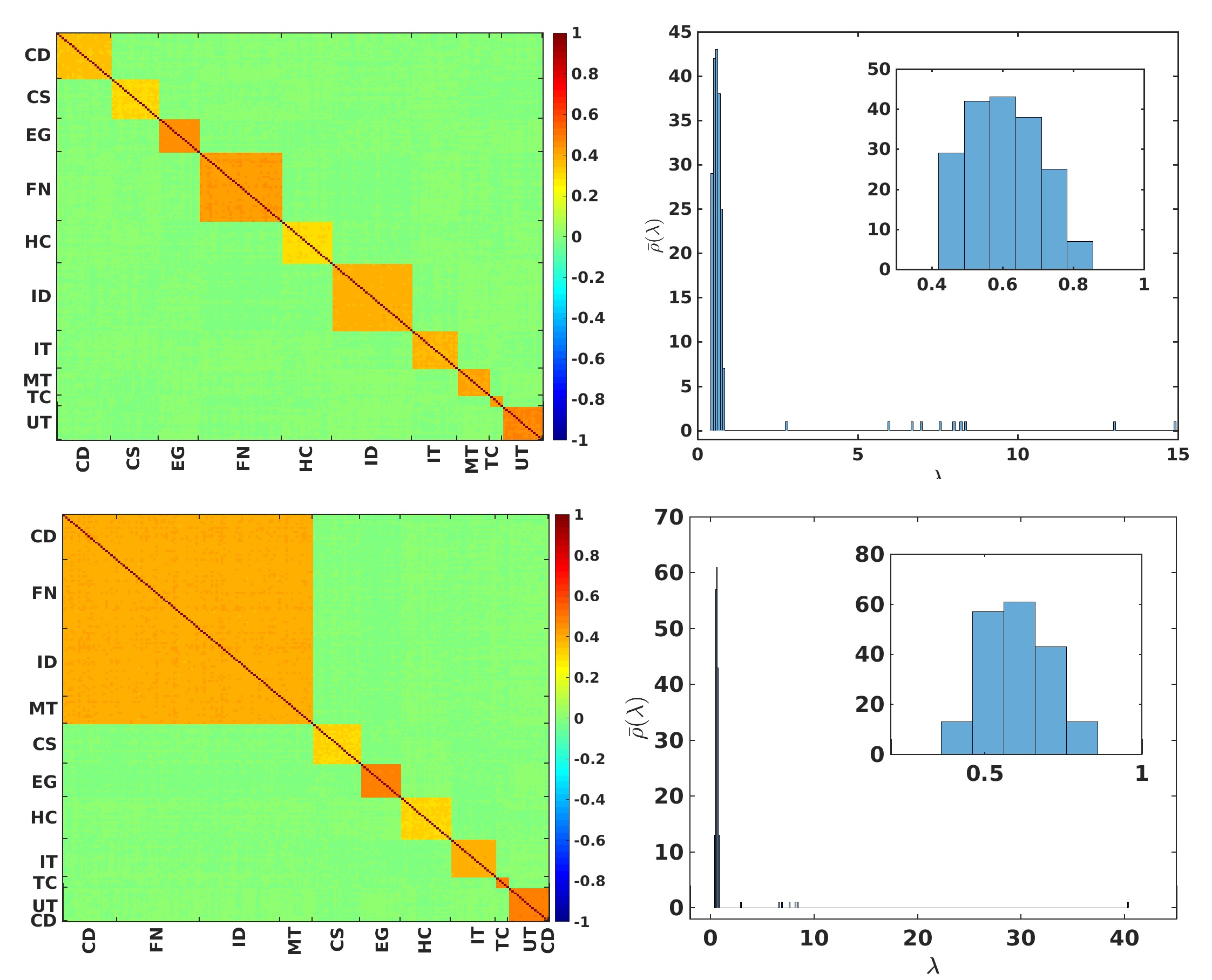}
\llap{\parbox[b]{2.8in}{(\textbf{a})\\\rule{0ex}{2.1in}}}
\llap{\parbox[b]{1.4in}{(\textbf{b})\\\rule{0ex}{2.1in}}}
\llap{\parbox[b]{2.9in}{(\textbf{d})\\\rule{0ex}{1in}}}
\llap{\parbox[b]{1.5in}{(\textbf{e})\\\rule{0ex}{1in}}}
\includegraphics[width=0.35\linewidth]{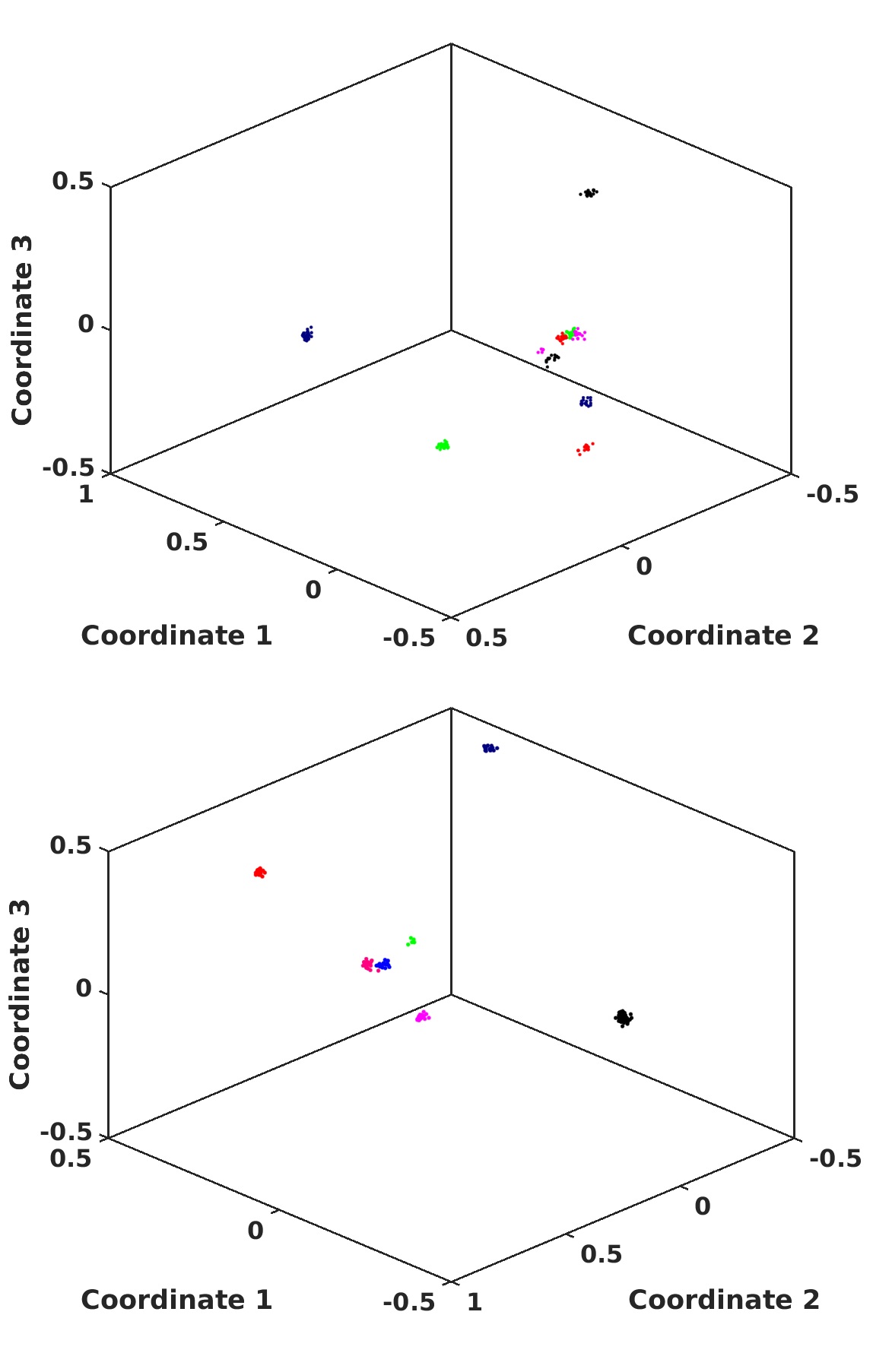}
\llap{\parbox[b]{1.7in}{(\textbf{c})\\\rule{0ex}{2.1in}}}
\llap{\parbox[b]{1.7in}{(\textbf{f})\\\rule{0ex}{1in}}}
\caption{(a) Cross-correlation matrices constructed from the correlated Gaussian time series with $10$ diagonal blocks of different correlations (equal to the average correlation of each sector in Fig.~\ref{fig_USA_avg_corr} (a)). (d) shows the same cross-correlation matrix but with one big block and $6$ smaller blocks. The mean correlation of the big block  is equal to the mean correlation of four sectors (\textbf{CD}, \textbf{FN}, \textbf{ID} and \textbf{MT} of Fig.~\ref{fig_USA_avg_corr} (a)). They have high inter-sectorial correlation over the last 32 years in S\&P 500 market. (b) and (e) show the eigenvalue spectra of the correlation matrices, which consist of the Mar\u{c}enko-Pastur distributions followed by $10$ group modes corresponding to $10$ sectors and $7$ group modes corresponding to $7$ sectors,  respectively.  Insets show the enlarged pictures of the random part of the spectrum. (c) and (f) show plots of $10$ and $7$ different clusters, respectively, drawn in different colors using $3$-dimensional $k$-means clustering technique. The clustering was performed on $3$-$D$ multidimensional scaling (MDS) map of $194$ stocks. Each point on the MDS map represents a stock of the market. The points are scattered in the map, based on the cross-correlations among the stocks -- more correlated stocks are placed nearby and less correlated are placed far apart (see also Ref.~\cite{Pharasi_2018}).} 
\label{fig_block_corr}
\end{figure}

Fig.~\ref{fig_block_corr} (a)  shows the cross-correlation matrices constructed from \textit{surrogate} data ($N=194$ correlated Gaussian noises, each of length $T=10000$) such that the matrix has $10$ diagonal blocks of different correlations (equal to the average correlations of different sectors of the  S\&P 500 market). Fig.~\ref{fig_block_corr} (d) shows the \textit{surrogate} cross-correlation matrix ($N=194; T=10000$) but now with one big block and $6$ smaller blocks. The mean correlation of the big block  is equal to the mean correlation of four sectors (\textbf{CD}, \textbf{FN}, \textbf{ID} and \textbf{MT} of Fig.~\ref{fig_USA_avg_corr} (a)) and they show high inter-sectorial correlation in S\&P 500 market in 32 years. 
Eigenvalue spectra of the correlation matrices are shown in Figs.~\ref{fig_block_corr} (b) and (e), each of which consists of the Mar\u{c}enko-Pastur distributions (see insets), followed by $10$ (and $7$) eigenvalues corresponding to $10$ (and $7$) blocks (similar to sectors), respectively.  Figs.~\ref{fig_block_corr} (c) and (f) show the $3D$ MDS plots, where the points (representing stocks) are scattered based on the correlations among the $10$ and $7$ blocks, respectively. In the MDS maps, more correlated stocks are placed nearby and anti-correlated are placed far apart (see also Ref.~\cite{Pharasi_2018}). The $k$-means clustering performed on the surrogate data matrices, with $k=10$ and $k=7$, yield $10$ and $7$ different clusters (represented in different colors), respectively. 

%%>>>>>>>>>>>>>>>>>>>>>>>>>>>>>>>>>>>>>>>>>>>>>>>>>>>>>>>>>>>>>>>>>>>
\subsubsection{Dynamics of the Correlation Structure of US market}
Next, we study the time evolution of the market correlations computed with the daily returns of $N=194$ stocks of S\&P 500 over the period of 32-year (1985-2016, with $T=8068$ trading days). 

%%%%%*********************************
\begin{figure}[!h]
\centering
\includegraphics[width=0.48\linewidth]{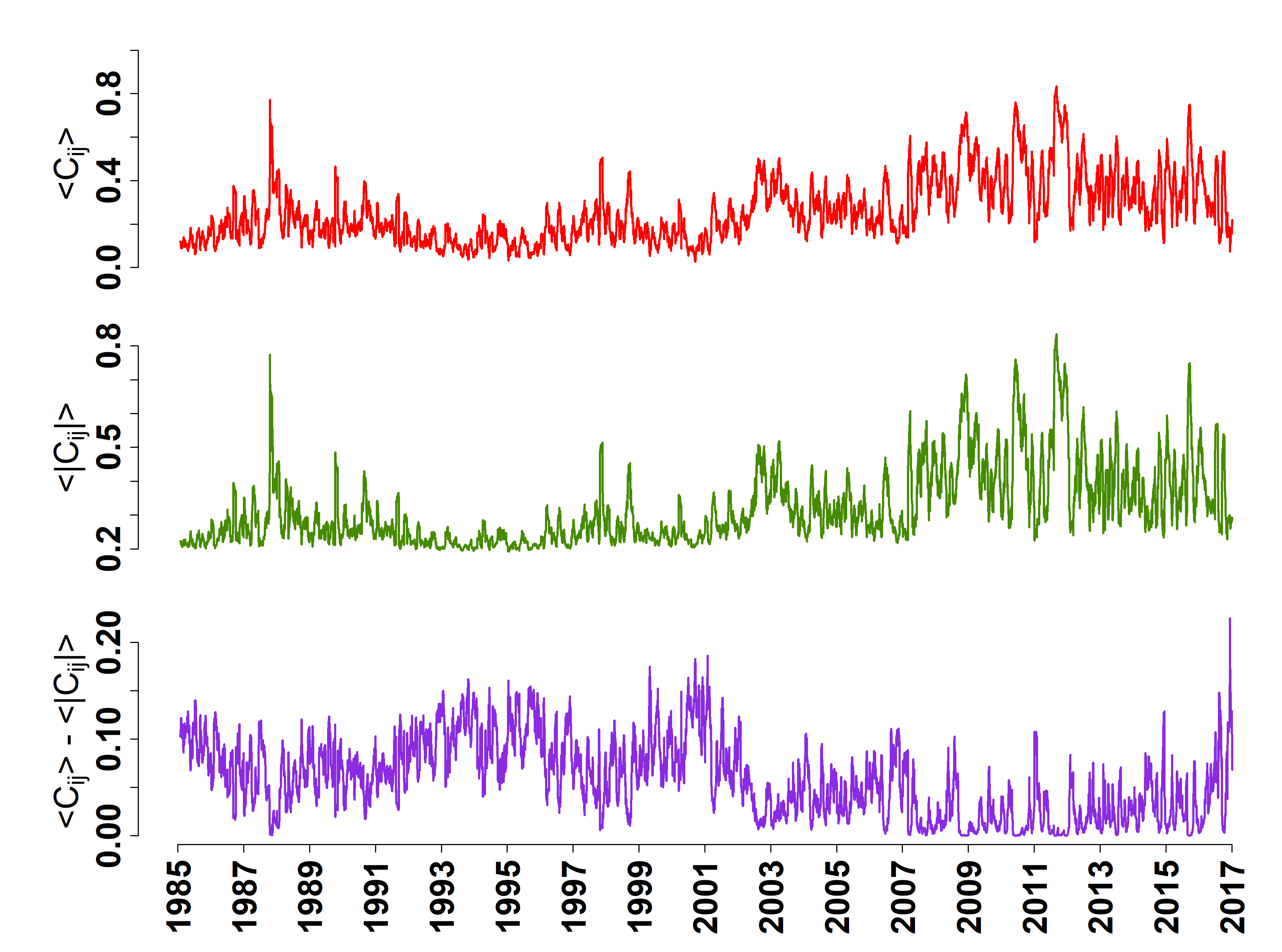}
\llap{\parbox[b]{2.3in}{(\textbf{a})\\\rule{0ex}{1.4in}}}
\includegraphics[width=0.48\linewidth]{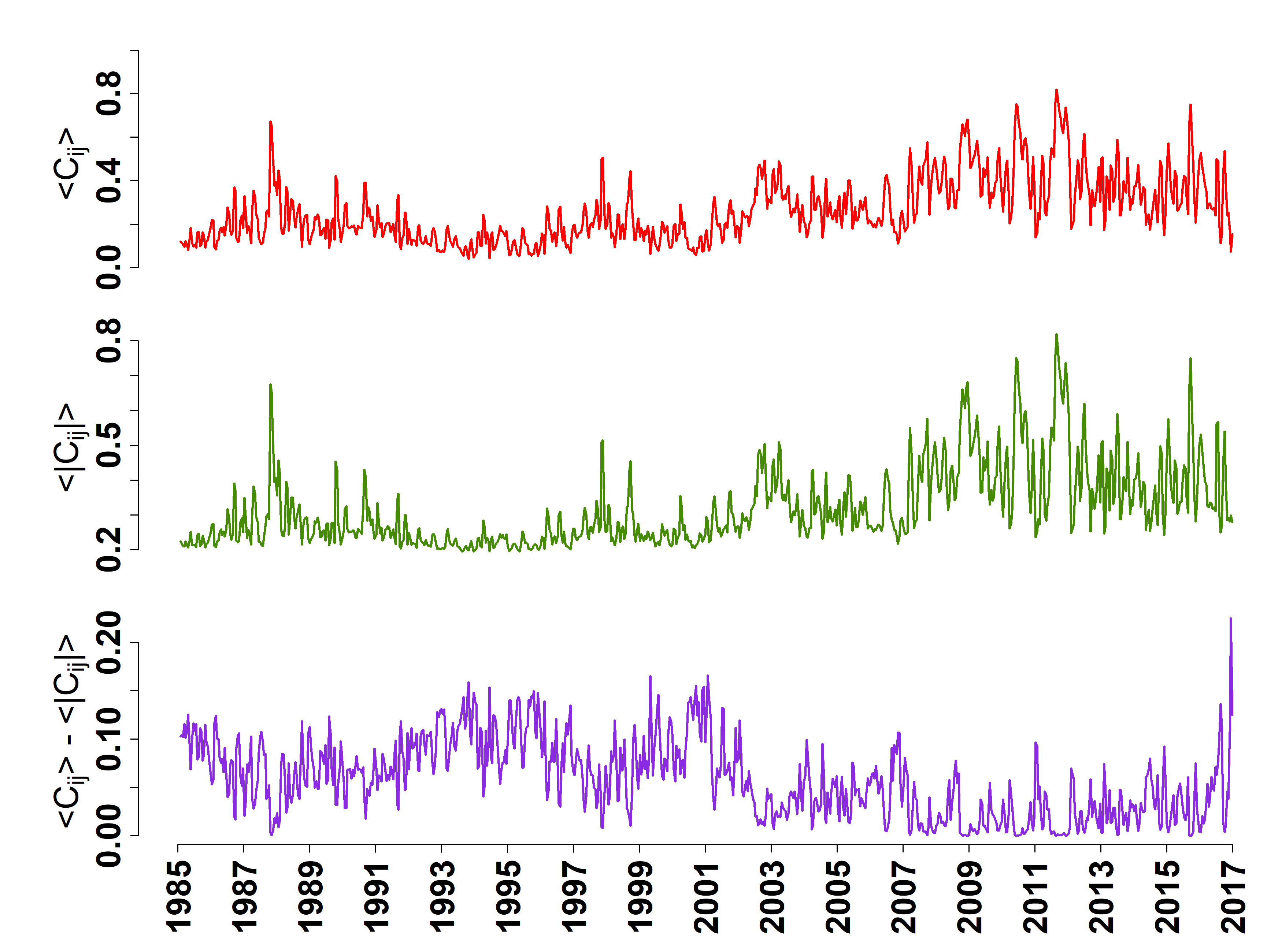}\llap{\parbox[b]{2.31in}{(\textbf{b})\\\rule{0ex}{1.4in}}}
\includegraphics[width=0.48\linewidth]{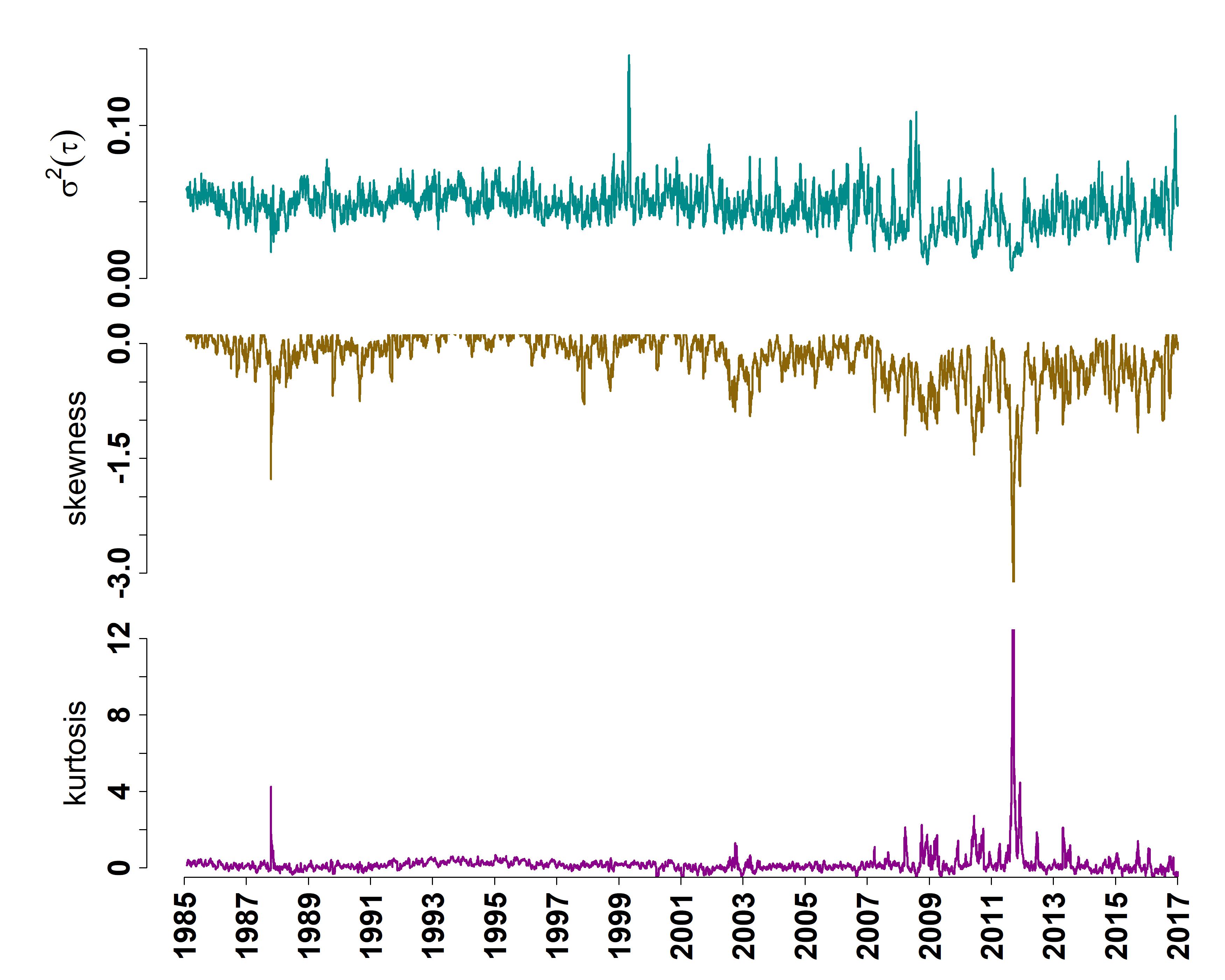}
\llap{\parbox[b]{2.3in}{(\textbf{c})\\\rule{0ex}{1.7in}}}
\includegraphics[width=0.48\linewidth]{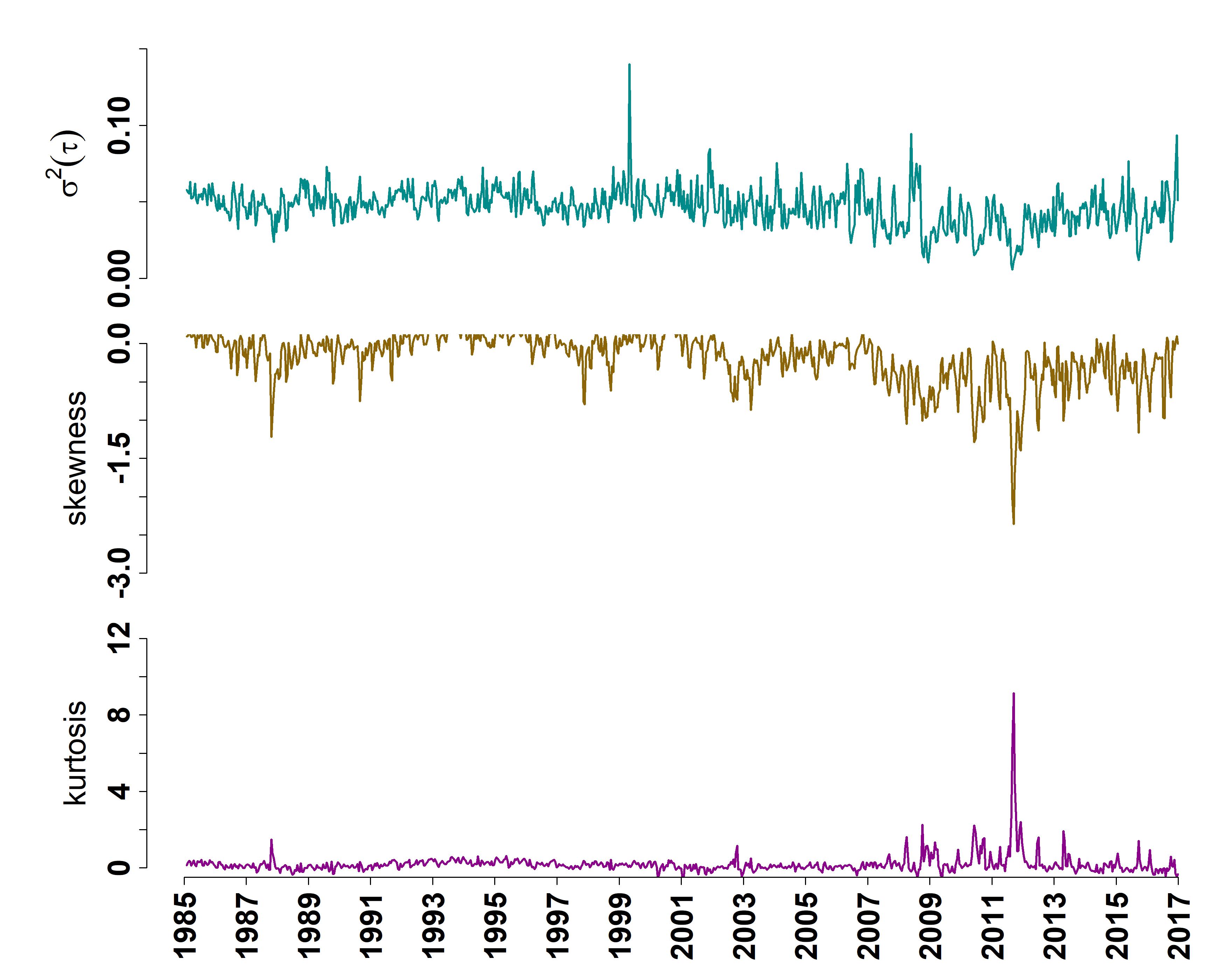}
\llap{\parbox[b]{2.3in}{(\textbf{d})\\\rule{0ex}{1.7in}}}
\caption{Plots of mean of correlation coefficients ($<C_{ij}>$), mean of absolute values of correlation coefficients ($<|C_{ij}|>$) and the difference $(df=<|C_{ij}|>-<C_{ij}>)$ as functions of time, for short epochs of $M=20$ days, and shifts of: (a) $\Delta \tau =1$ day and (b) $\Delta \tau =10$ days.  We find that during crashes (when mean correlation is very high), the difference $df=<|C_{ij}|>-<C_{ij}>$ shows a minimum (close to zero) (see Ref.~\cite{Mijail_2018}). Plots of variance ($\sigma^2$), skewness, and kurtosis of the correlation coefficients as functions of time, for short epochs of $M=20$ days, and shifts of: (c) $\Delta \tau =1$ day and (d) $\Delta \tau =10$ days.}
\label{fig_moments}
\end{figure}
%%%%%*********************************
%%%%%*********************************
\begin{figure}[!h]
\centering
\includegraphics[width=0.95\linewidth]{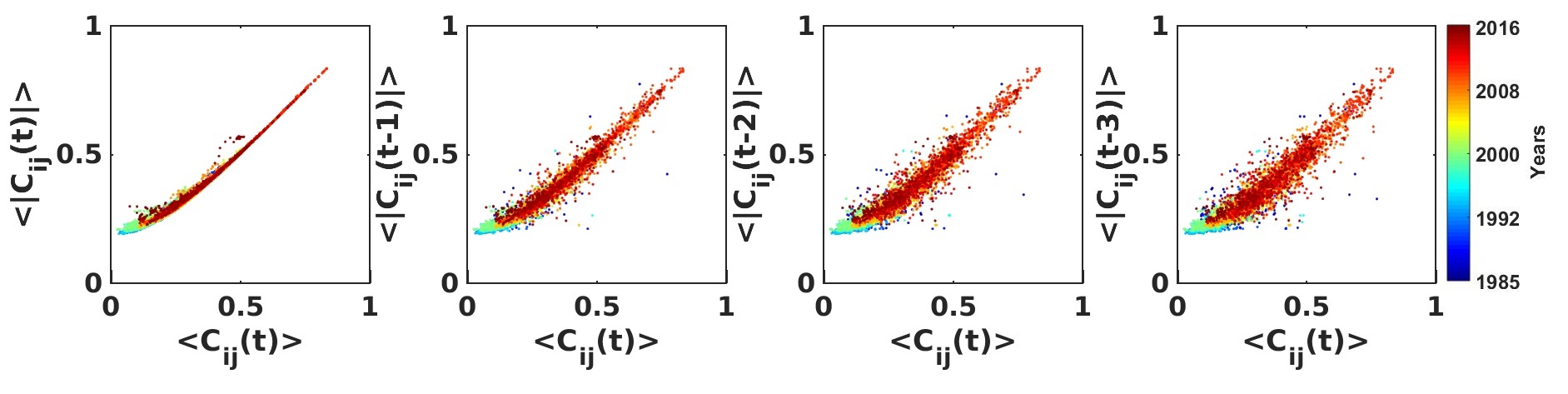}
\llap{\parbox[b]{4.4in}{(\textbf{a})\\\rule{0ex}{1.1in}}}
\includegraphics[width=0.95\linewidth]{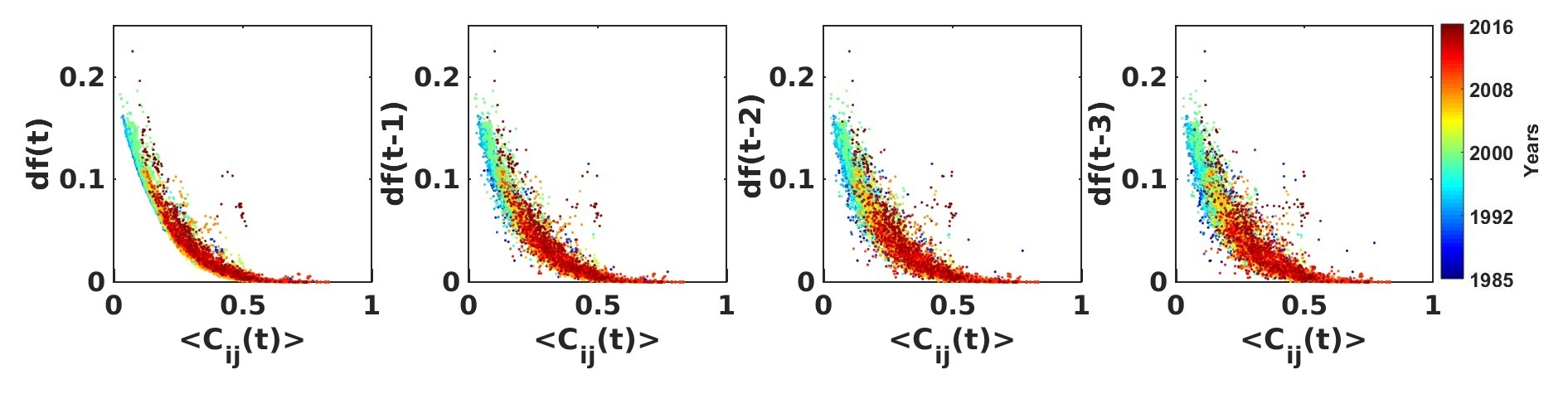}
\llap{\parbox[b]{4.4in}{(\textbf{b})\\\rule{0ex}{1.1in}}}
\includegraphics[width=0.95\linewidth]{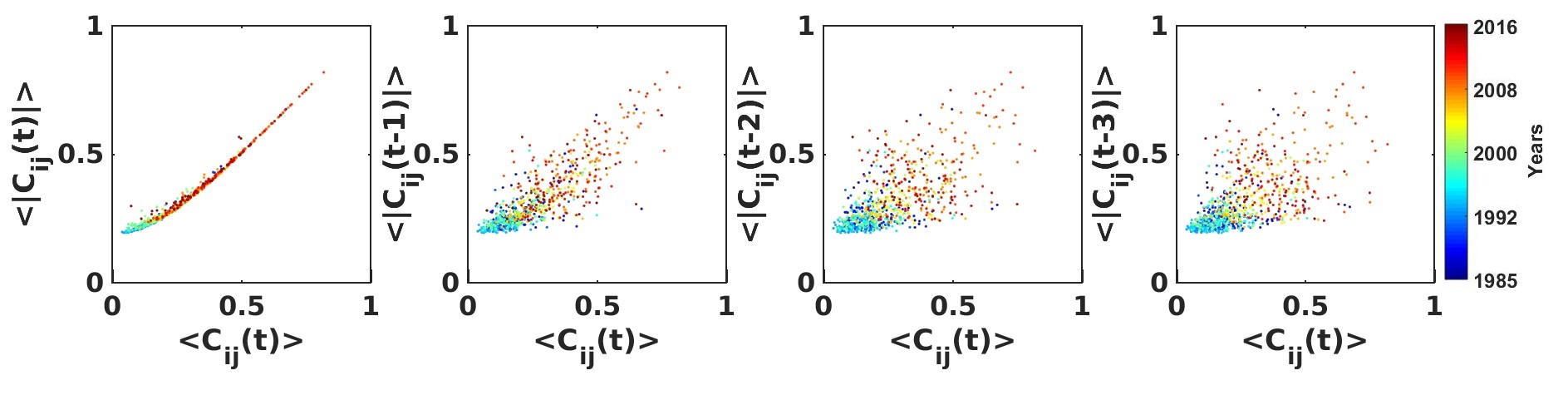}
\llap{\parbox[b]{4.4in}{(\textbf{c})\\\rule{0ex}{1.1in}}}
\includegraphics[width=0.95\linewidth]{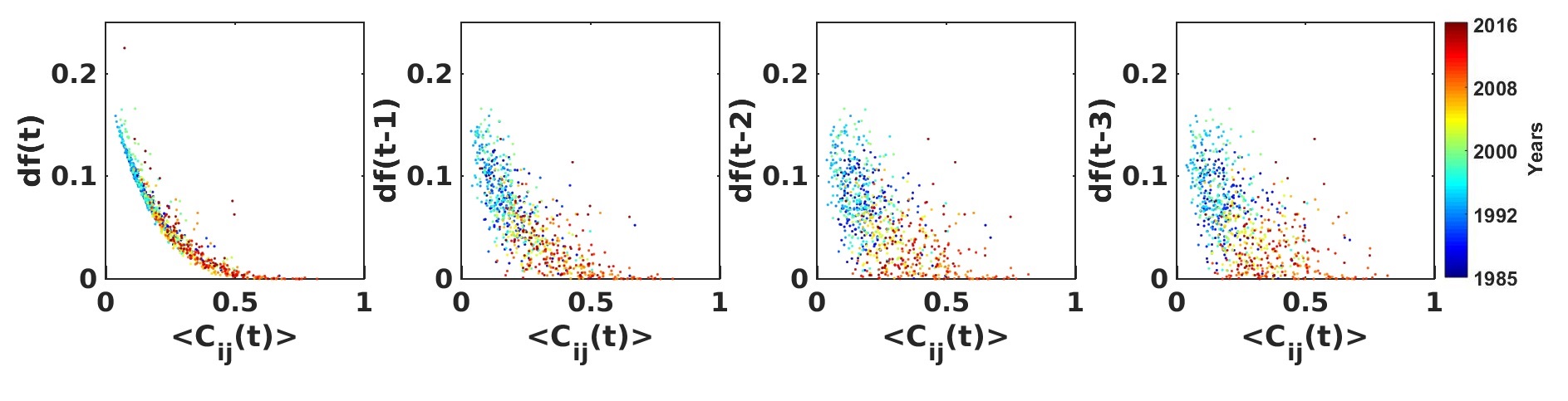}
\llap{\parbox[b]{4.4in}{(\textbf{d})\\\rule{0ex}{1.1in}}}
\caption{Scatter plots of $<C_{ij}>$ vs. $<|C_{ij}|>$ and  $<C_{ij}>$ and $df=<|C_{ij}|>-<C_{ij}>$, for different time lags (No lag,1-day, 2-days and 3-days) for the correlation matrix of epoch $20$ days, with shifts of: (a)-(b)  $\Delta \tau =1$ day; (c)-(d) $\Delta \tau =10$ days. The color-bar shows the time period in years.} 
\label{fig_scatter}
\end{figure}
%%%%%*********************************
Fig.~\ref{fig_moments} (a) and (b) show plots of mean of correlation coefficients ($<C_{ij}>$), mean of absolute values of correlation coefficients ($<|C_{ij}|>$) and the difference of the mean and the mean of absolute values of correlation coefficients $df=<|C_{ij}|> - <C_{ij}>$ for short epochs of $M = 20$ days, with shifts of: $\Delta \tau=1$ day  ($95\%$ overlap) and $\Delta \tau=10$ days ($50\%$ overlap), respectively. Shifts toward the positive side of correlations are pointing toward periods of market crashes (with very high mean correlation values). The values of $df$ are anti-correlated with the values of the mean of correlation coefficients.  During a market crash when mean of correlation coefficient is high, there are very little anti-correlations among the stocks, then the value of $df$ is extremely small, indeed near to zero (see Ref.~\cite{Mijail_2018}). It may act as an indicator of a market crash, as we observe that there is a high anti-correlation between the values of $df$ and $<C_{ij}>$, with leads of one or two days (ahead of the market crashes). Similarly, Fig.~\ref{fig_moments} (c) and (d) show the plots of variance, skewness, and kurtosis of the correlation coefficients $C_{ij}$ as functions of time with shifts of  $\Delta \tau=1$ day and  $\Delta \tau=10$ days, respectively. The mean correlation is anti-correlated to variance and skewness of $\boldsymbol C$, i.e., when the mean correlation is high then both variance and skewness are low. Kurtosis is highly correlated to the mean correlation. These observations are seen in the dynamical evolution of the market with  epochs of $M=20$ days, and shifts of $\Delta \tau=1,10$ day(s).

%%%%%*********************************
\begin{figure}[!h]
\centering
\includegraphics[width=0.49\linewidth]{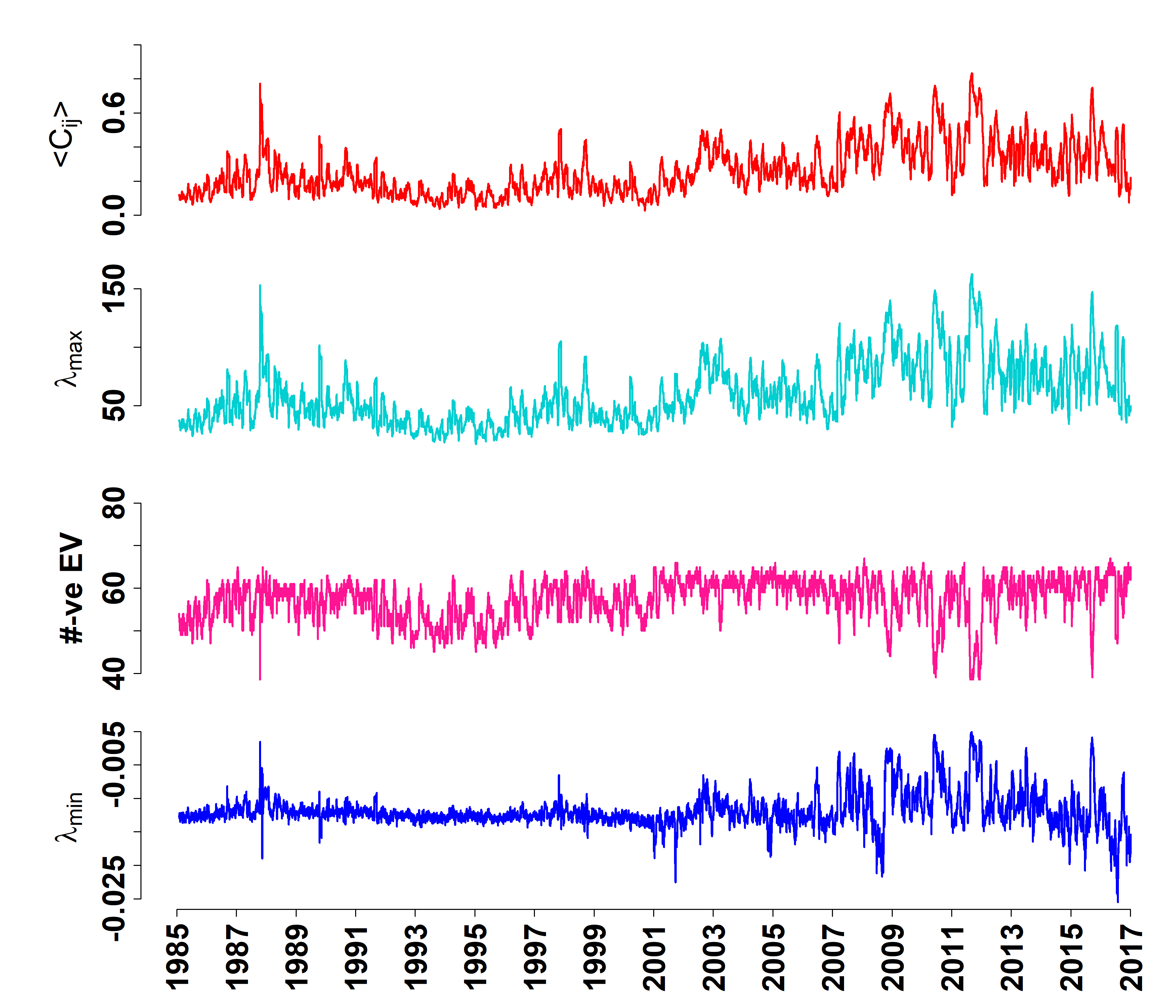}
\llap{\parbox[b]{2.3in}{(\textbf{a})\\\rule{0ex}{1.8in}}}
\includegraphics[width=0.49\linewidth]{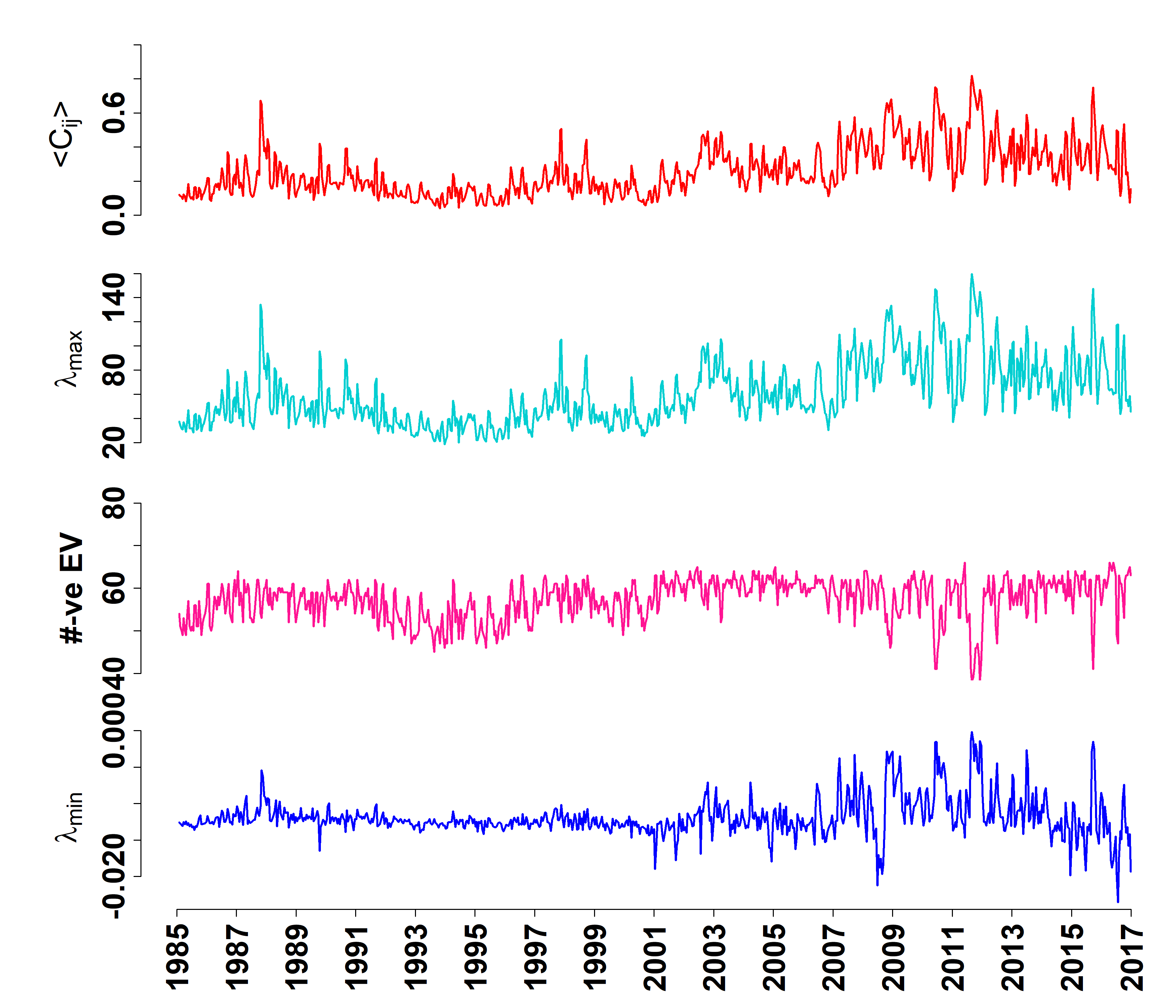}
\llap{\parbox[b]{2.3in}{(\textbf{b})\\\rule{0ex}{1.8in}}}
\caption{Plots for mean of correlation coefficients ($<C_{ij}>$), maximum eigenvalue ($\lambda_{max}$), number of negative eigenvalues ($\#-ve~EV$) and smallest eigenvalue ($\lambda_{min}$) of the spectrum as a function of time for an epoch of $20$ days at $\epsilon=0.01$ with shifts of: (a) $\Delta \tau =1$ day and (b) $\Delta \tau =10$ days. The correlation between $<C_{ij}>$ and $\lambda_{max}$ is high, but two other properties of the ``emerging spectrum" ($\#-ve~EV$ and $\lambda_{min}$ ) are less correlated to mean correlation $<C_{ij}>$.}
\label{fig_NEV}
\end{figure}
%%%%%*********************************

The scatter plots between $<C_{ij}>$ and $<|C_{ij}|>$, and $<C_{ij}>$ and $df (=<|C_{ij}|>-<C_{ij}>)$ for different time lags (no-lag, lag-1, lag-2, and lag-3) of empirical correlation matrices $\boldsymbol C(\tau)$, with $194$ stocks of S\&P 500 and epochs of $M=20$ days, and shift of $\Delta \tau =1$ day, are shown in Fig.~\ref{fig_scatter} (a) and (b), respectively. Here lag-1, lag-2, and lag-3 represent time lags of  1 day, 2 days, and 3 days, respectively. The color-bar shows the time period from 1985 to 2016 in years. The scatter plots show the correlations among  $<C_{ij}>$ vs. $<|C_{ij}|>$ and $<C_{ij}>$  vs. $df$, at different time lags. The variances of the scatter plots increase with the increase of time lag, keeping the value of linear correlation coefficient almost similar. The strong linear correlation between $<C_{ij}>$ and $<|C_{ij}|>$ may give us an early information about a crash up to 3 days ahead (from the result of lag-3). Similar linear correlations are also visible in Fig.~\ref{fig_scatter} (c) and (d), between $<C_{ij}>$ and $<|C_{ij}|>$, and $<C_{ij}>$ and $df$, at different time lags (no-lag, lag-1, lag-2, and lag-3) for a shift of $\Delta \tau =10$  days. Here, obviously lag-1, lag-2, and lag-3 represent time lags of  10 days, 20 days, and 30 days, respectively. The large variances in scatter plots indicate that it is hard to detect and extract information about a crash, e.g., 30 days in advance. 

Fig.~\ref{fig_NEV} (a) shows the temporal variation of mean correlation ($<C_{ij}>$), maximum eigenvalue ($\lambda_{max}$), number of negative eigenvalues ($\#-ve~EV$) and smallest eigenvalue ($\lambda_{min}$) of the emerging spectra with a shift of $\Delta \tau =1$ day. Using a small distortion ($\epsilon=0.01$), we break the degeneracy of eigenvalues at zero and get the ``emerging spectra" of eigenvalues which contain some interesting infromation about the market.  The effect of the small distortion parameter $\epsilon=0.01$ is negligible on non-zero eigenvalues of the spectrum including $\lambda_{max}$. We high correlation between $<C_{ij}>$ and $\lambda_{max}$. But the other properties of emerging spectrum ($\#-ve~EV$ and $\lambda_{min}$) are less correlated with mean correlation$<C_{ij}>$ \cite{Ochoa_2018}. Fig.~\ref{fig_NEV} (b) shows the same for the shift of $\Delta \tau =10$ days.
%>>>>>>>>>>>>>>>>>>>>>>>>>>>>>>>>>>>>>>>>>>>>>>>>>>>>>>>>>>>>>>>>>>>>>>>>>>>>>>>>>>>>>>>>>>>
\section{Recent applications of RMT in financial markets}
\subsection{Identification of Market states and long-term precursors to a crash state}

%%%%%*********************************
\begin{figure}[!h]
\centering
\includegraphics[width=0.55\linewidth]{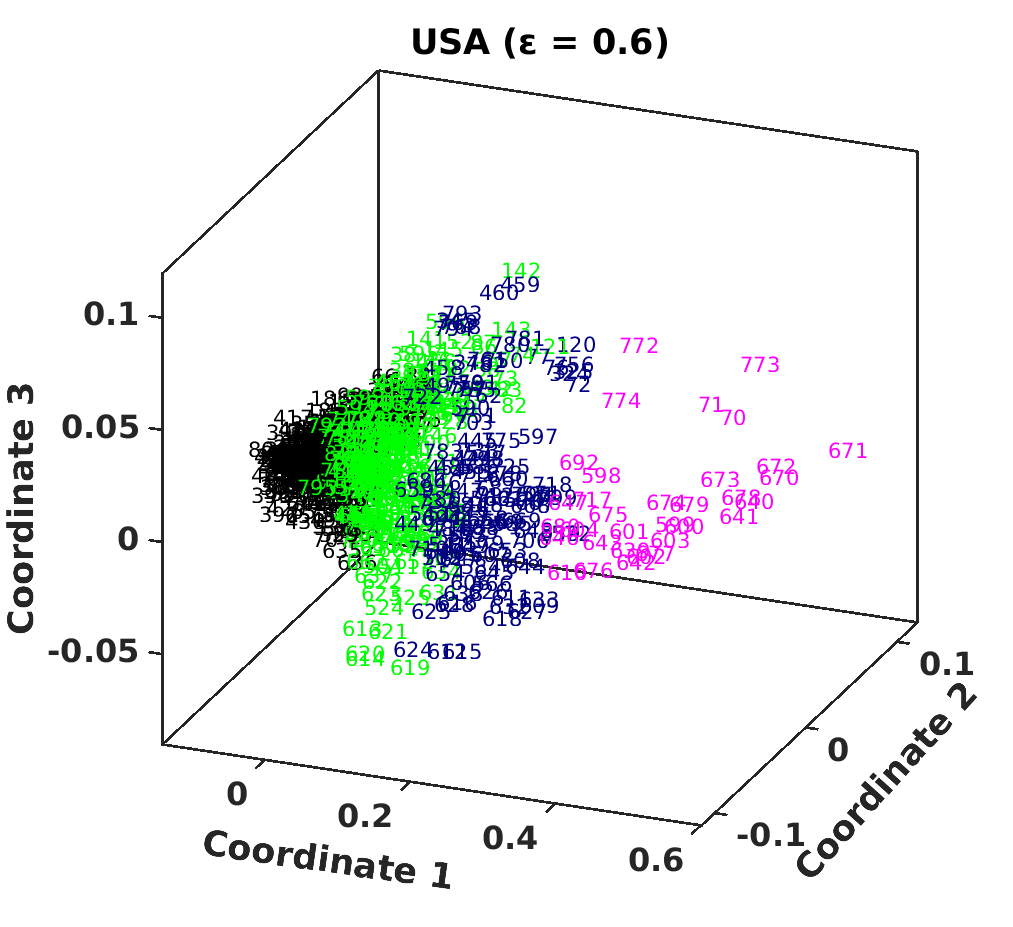}\llap{\parbox[b]{2.3in}{(\textbf{a})\\\rule{0ex}{1.9in}}}\includegraphics[width=0.45\linewidth]{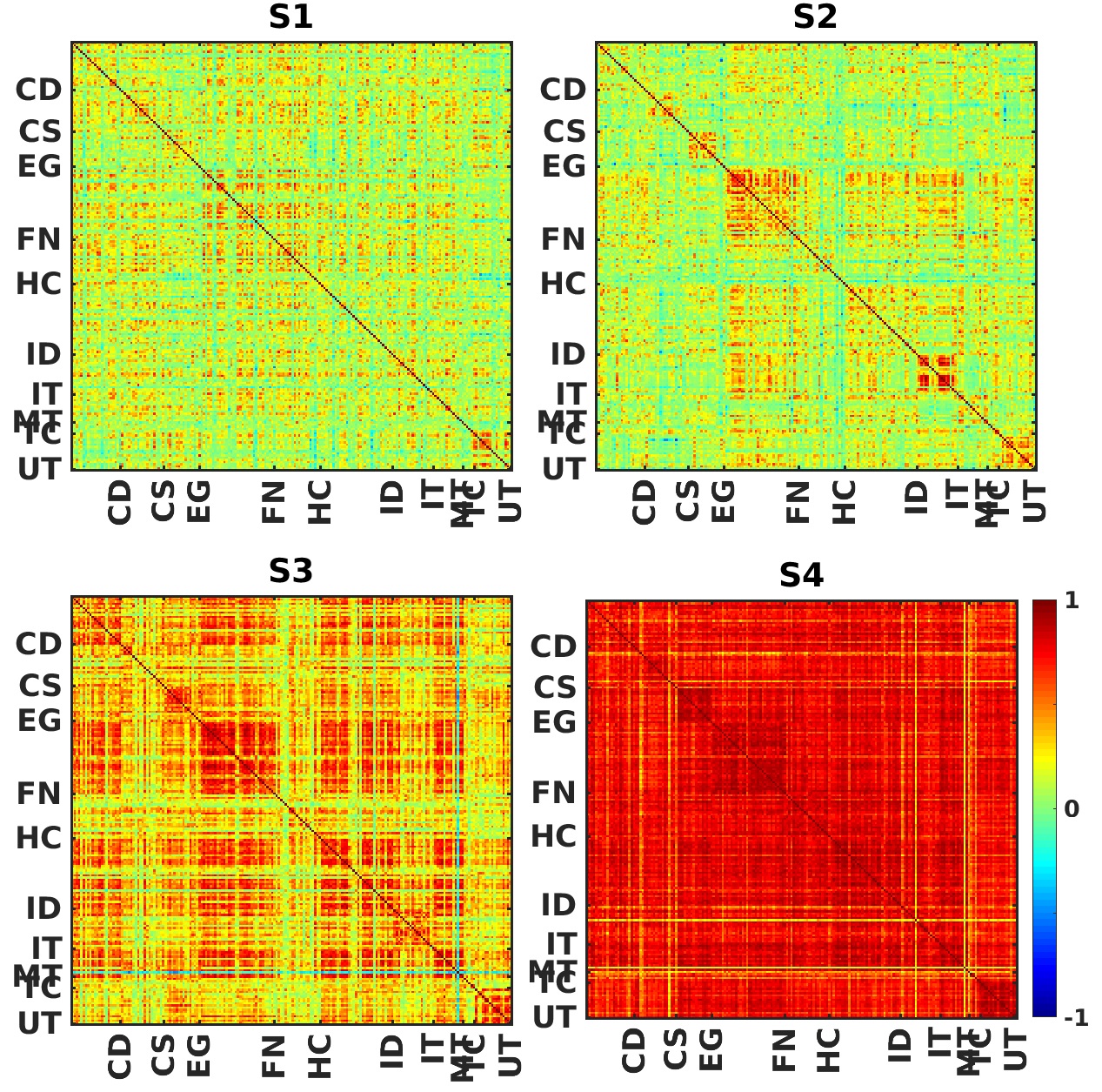}\llap{\parbox[b]{2.3in}{(\textbf{b})\\\rule{0ex}{1.9in}}}\\\includegraphics[width=8cm,height=2.5cm]{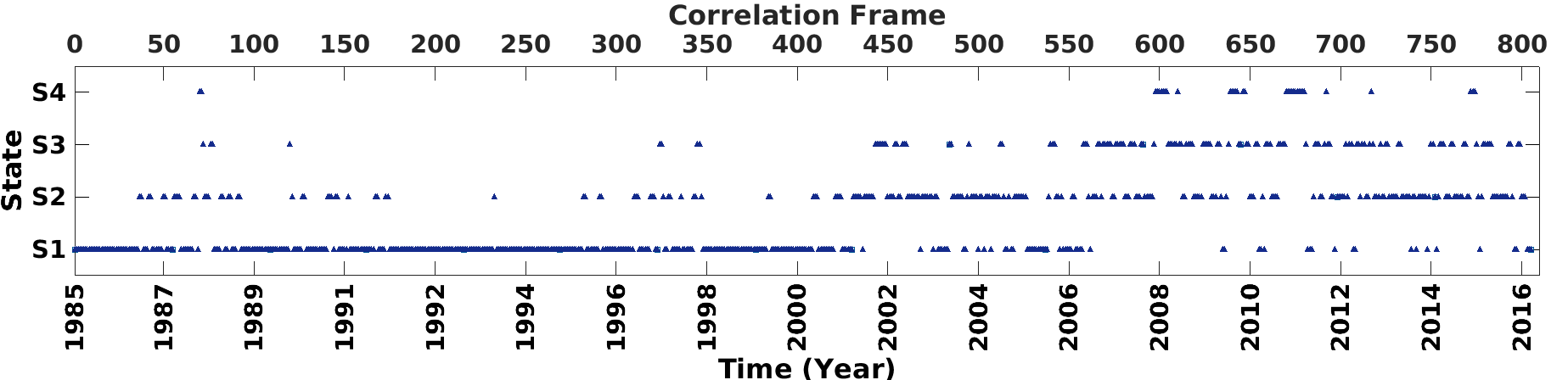}\llap{\parbox[b]{3.2in}{(\textbf{c})\\\rule{0ex}{.9in}}}\includegraphics[width=0.33\linewidth]{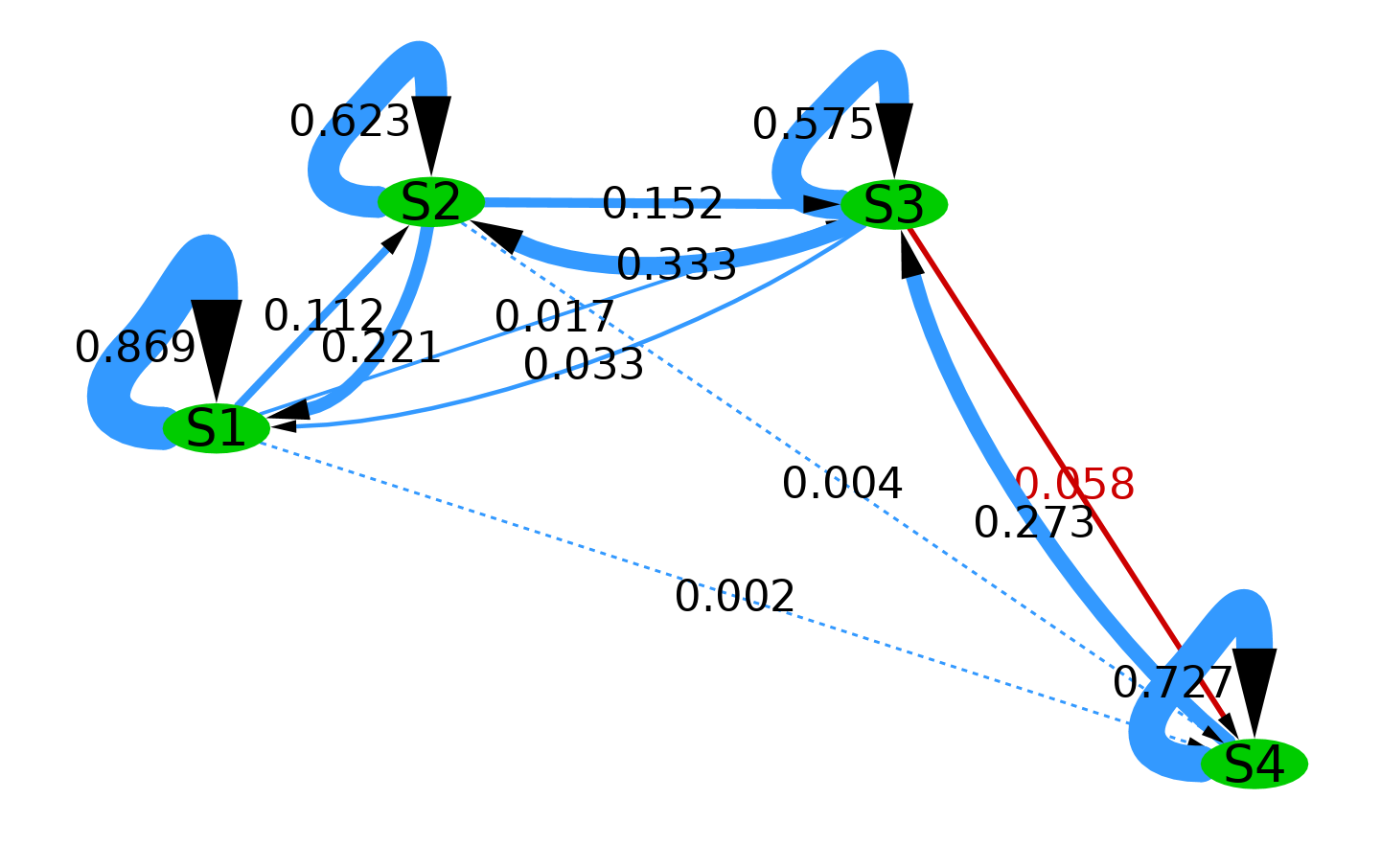}\llap{\parbox[b]{1.5in}{(\textbf{d})\\\rule{0ex}{.9in}}}
\caption{(a) Classification of the US market into four typical market states. $k$-means clustering is performed on a MDS map constructed from noise suppressed ($\epsilon=0.6$) similarity matrix \cite{Munnix_2012}. The coordinates assigned in the MDS map are the corresponding correlation matrices constructed from short-time series of $M=20$ days and shifted by $\Delta\tau =10$ days. (b) shows the four different states of the US market as S1, S2, S3 and S4, where S1 corresponds to a calm state with low mean correlation, and S4 corresponds to a critical state (crash) with high mean correlation. (c) Temporal dynamics of the US market in four different states ($S1, S2, S3$ and $S4$) for the period of $1985-2016$. (d) Network plot for transition probabilities of paired market states (MS). The transition probability of paired market states going from $S1$ and $S2$ to $S4$ is much lesser than $1\%$ but from $S3$ to $S4$ is $6\%$. Figure adapted from Ref.~\cite{Pharasi_2018}.} 
\label{fig_Appl}
\end{figure}
%%%%%*********************************
The study of the critical dynamics in any complex system is interesting, yet it can be very challenging. Recently, Pharasi et al. \cite{Pharasi_2018} presented an analysis based on the correlation structure patterns of S\&P 500 (USA) data and Nikkei 225 (JPN) data, with short time epochs during the 32-year period of 1985-2016. They identified ``market states'' as clusters of similar correlation structures which occurred more frequently than by pure chance (randomness). 

They first used the power mapping to reduce noise of the singular correlation matrices and obtained distinct and denser clusters in three dimensional MDS map (as shown in Fig.~\ref{fig_Appl}(a)). The effects of noise-suppression were found to be prominent not only on a single correlation matrix at one epoch, but also on the similarity matrices computed for different correlation matrices at different short-time epochs, and their corresponding MDS maps. 
Using 3D-multidimensional scaling maps, they applied $k$-means clustering to divide the clusters of similar correlation patterns into $k$ groups or market states. One major difficulty of this clustering method is that one has to pass the value of $k$ as an input to the algorithm.  Normally, there are several proposed methods of determining the value of $k$ (often arbitrary). Pharasi et al. \cite{Pharasi_2018} showed that using a new prescription based on the cluster radii and an optional choice of the noise suppression parameter, one could have a fairly robust determination of the ``optimal'' number of clusters. 
 
In the new prescription, they measured the mean and the standard deviation of the intra-cluster distances using an ensemble of fairly large number (about 500) of different initial conditions (choices of random coordinates for the $k$-centroids or equivalently random initial clustering of $n$ objects); each set of initial conditions usually results in slightly different clustering of the $n$ objects representing different correlation matrices. If the clusters of points are very distinct in the coordinate space, then even for different initial conditions, the $k$-means clustering method yields same results, producing a small variance of the intra-cluster distance. However, the problem of allocating the matrices into the different clusters becomes problematic, when the clusters are very close or overlapping, as the initial conditions can then influence the final clustering of the different points; so there is a larger variance of the intra-cluster distance for the ensemble of initial conditions. Therefore, a minimum variance or standard deviation for a particular number of clusters implies the robustness of the clustering.
For optimizing the number of clusters, Pharasi et al. proposed that one should look for \textit{maximum} $k$, which has the \textit{minimum variance} or standard deviation in the intra-cluster distances with different initial conditions. 
Thus, based on the modified prescription of finding similar clusters of correlation patterns, they characterized the market states for USA and JPN.
 
Here, in Fig.~\ref{fig_Appl}(b), we reproduce the results for the US market, showing four typical market states. The evolution of the market can be then viewed as the dynamical transitions between market states, as shown in Fig.~\ref{fig_Appl}(c). Importantly, this method yields the correlation matrices that correspond to the critical states (or crashes). They correspond to the well-known financial market crashes and clustered in market state $S4$. They also analyzed the transition probabilities of the paired  market states, and found that (i) the probability of remaining in the same state is much higher than the transition to a different states, and (ii) most probable transitions are the nearest neighbor transitions, and the transitions to other remote states are rare (see Fig.~\ref{fig_Appl}(d)). Most significantly, the state adjacent to a critical state (crash) behaved like a long-term ``precursor'' for a critical state, serving an early warning for a financial market crash.

%>>>>>>>>>>>>>>>>>>>>>>>>>>>>>>>>>>>>>>>>>>>>>>>>>>>>>>>>>>>>>>>>>>>>>>>>>>>>>>>>>>>>>>>>>>
\subsection{Characterization of catastrophic instabilities}
Market crashes, floods, earthquakes, and other catastrophic events, though rarely occurring, can have devastating effects with long term repurcussions. Therefore, it is of primal importance to study the complexity of the underlying dynamics and signatures of catastrophic events. Recently, Sharma et al. \cite{Chakraborti_2018} studied the evolution of cross-correlation structures of stock return matrices and their eigenspectra over different short-time epochs for the US market and Japanese market. By using the power mapping method, they applied the non-linear distortion with a small value of distortion parameter $\epsilon=0.01$ to correlation matrices computed for any epoch, leading to the {\it emerging spectrum} of eigenvalues. 

\begin{figure}[!h]
\centering
\includegraphics[width=0.49\linewidth]{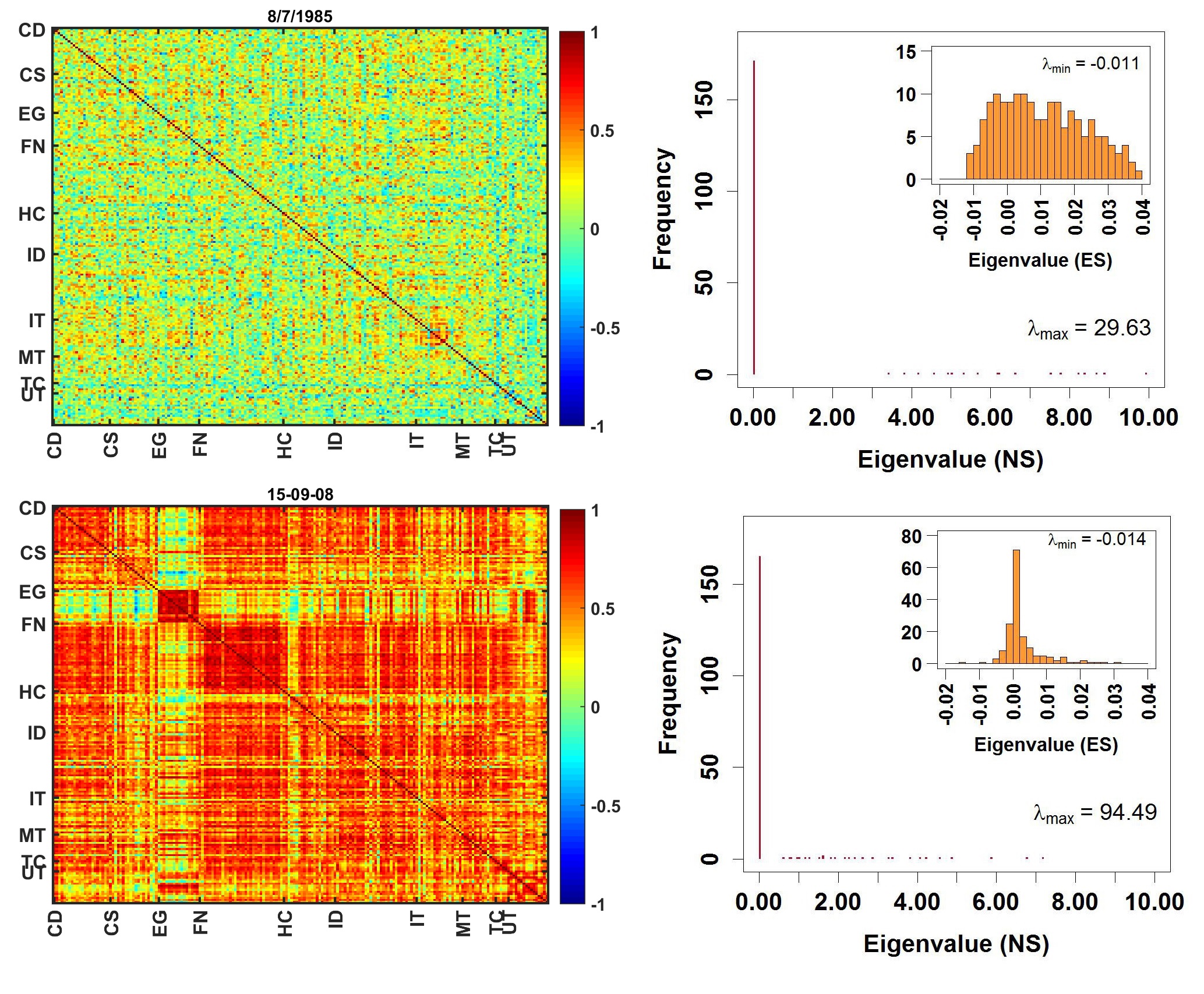}\llap{\parbox[b]{2.4in}{(\textbf{a})\\\rule{0ex}{1.8in}}}
\llap{\parbox[b]{2.4in}{(\textbf{b})\\\rule{0ex}{0.9in}}}
\includegraphics[width=0.48\linewidth]{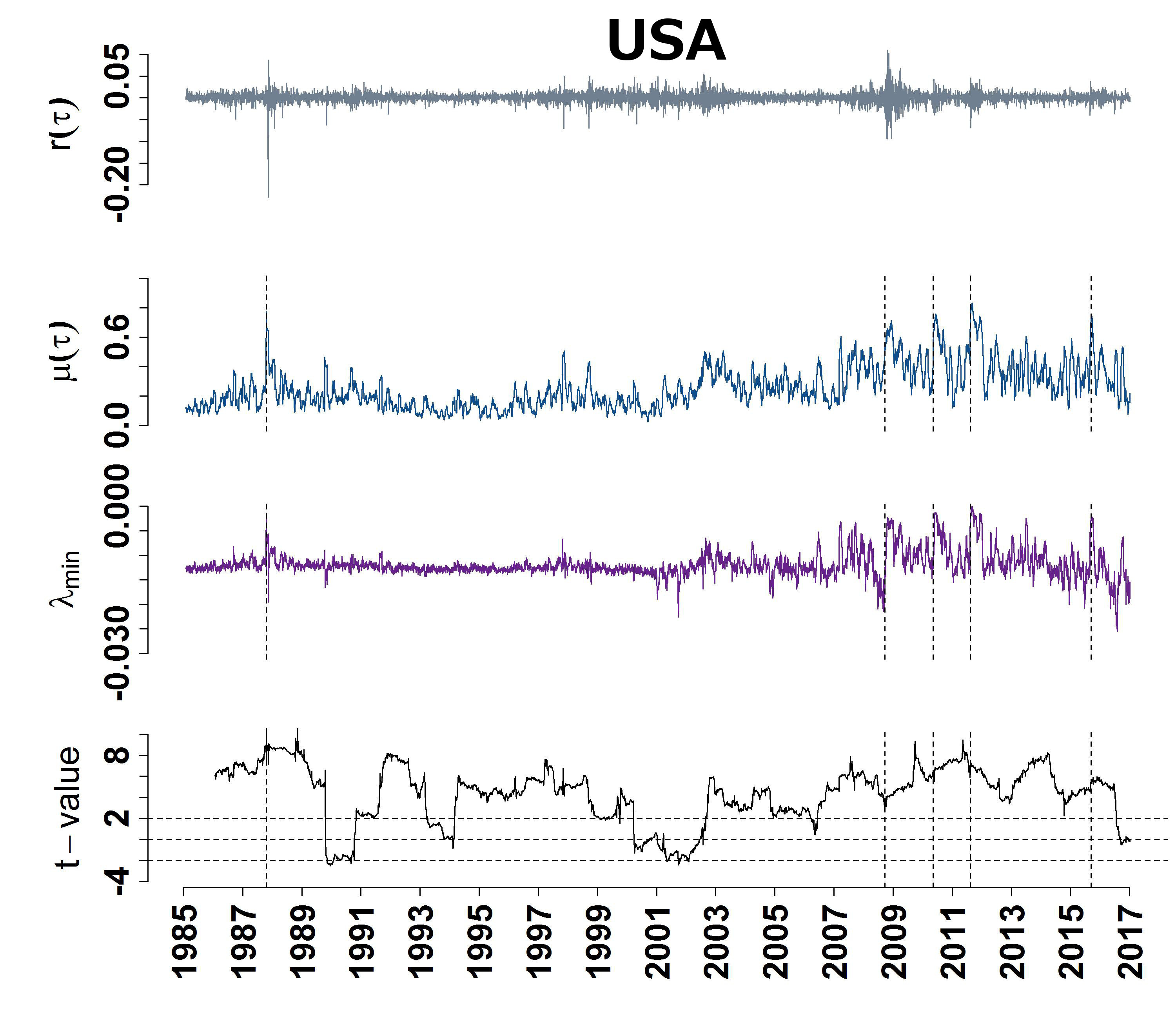}\llap{\parbox[b]{2.2in}{(\textbf{c})\\\rule{0ex}{1.8in}}}
\caption{(a) Non-critical (normal) period of the correlation matrix and its eigenvalue spectrum, evaluated for the \textit{short} return time series for an epoch of $M=20$ days ending on 08-07-1985, with the maximum eigenvalue of the normal spectrum $\lambda_{max}= 29.63$. Inset: Emerging spectrum using power map technique ($\epsilon= 0.01$) is a deformed semi-circle, with the smallest eigenvalue of the emerging spectrum $\lambda_{min}= -0.011$. (b) Critical (crash) period of the correlation matrix and its  eigenspectrum, evaluated for an epoch of $M=20$ days ending on 15-09-2008, with the maximum eigenvalue of the normal spectrum  $\lambda_{max}= 94.49$. Inset: Emerging spectrum using power map technique ($\epsilon= 0.01$) is Lorentzian, with the smallest eigenvalue of the emerging spectrum $\lambda_{min}= -0.014$. (c) USA (i) market return $r(t)$, (ii) mean market correlation $\mu (t)$, (iii) smallest eigenvalue of the emerging spectrum ($\lambda_{min}$), and (iv) t-value of the t-test, which tests the statistical effect over the lag-1 smallest eigenvalue $\lambda_{min}(t-1)$  on the mean market correlation $\mu(t)$. The mean of the correlation coefficients and the smallest eigenvalue in the emerging spectra are correlated to a large extent. Notably, the smallest eigenvalue behaves differently (sharply rising \textit{or} falling) at the same time when the mean market correlation is very high (crash). The vertical dashed lines correspond to the major crashes, which brewed due to internal market reactions. Note that, the smallest eigenvalue of the US market indicates that the financial market has become more turbulent from 2001 onward. Figure adapted from Ref.~\cite{Chakraborti_2018}.} 
\label{fig_App2}
\end{figure}

Here, we reproduce some of the significant findings of the paper \cite{Chakraborti_2018}. Interestingly, it is found that the statistical properties of the emerging spectrum display the following features: (i) the shape of the emerging spectrum reflects the market instability (see Fig.~\ref{fig_App2}(a) and (b)), (ii) the smallest eigenvalue (in a similar way as the maximum eigenvalue, which captured the mean correlation of the market) indicated that the financial market had become more turbulent, especially from 2001 on wards (see Fig.~\ref{fig_App2}(c)), and (iii) the smallest eigenvalue is able to statistically distinguish the nature of a market turbulence or crisis -- internal instability or external shock (see Fig.~\ref{fig_App2}(c)).  In certain instabilities the smallest eigenvalue of the emerging spectrum was positively correlated with the largest eigenvalue (and thus with the mean market correlation) while in other cases there were trivial anti-correlations. They proposed that this behavioral change could  be associated to the question whether a crash is associated to intrinsic market conditions (e.g., a bubble) or to external events (e.g., the Fukushima meltdown). A lead-lag effect of the crashes was also observed through the behavior of $\lambda_{min}$ and mean correlation, which could be examined further. 

%%%%%*********************************************************
%%%%%*********************************
%%%%**********************************
\section{Concluding Remarks}
We have presented a brief overview of the Wishart and correlated Wishart ensembles in the context of financial time series analysis. We displayed the dependence of the length of the time series on the eigenspectra of the Wishart ensemble. The eigenspectra of large random matrices are not very sensitive to $Q=T/N$; however, the amount of spurious correlations is dependent on it. To avoid the problem  of non-stationarity and suppress the noise in the correlation matrices, computed over short epochs, we applied the power mapping method on the correlation matrices.  We  showed that the shape of the emerging spectrum depends on the amount of the correlation $U$ of the correlated Wishart ensemble. We also studied the effect of the non-linear distortion parameter $\epsilon$ on the emerging spectrum.

Then we demonstrated the eigenvalue decomposition of the empirical cross-correlation matrix into market mode, group modes and random modes, using the return time series of $194$ stocks of S\&P 500 index during the period of 1985-2016. The bulk of the eigenvalues behave as random modes and give rise to the Mar\u{c}enko-Pastur. We also created surrogate correlation matrices to understand the effect of the sectoral correlations. Then we studied the eigenvalue distribution of those matrices as well as \textit{k}-means clustering on the MDS maps generated from the correlation matrices. Evidently, if we have $10$ diagonal blocks (representing sectors) then we get $10$ clusters on a MDS map. Similarly, when we merged the four blocks to one and had 7 diagonal blocks then again we got 7 clusters on the MDS map.

Further, we studied the dynamical evolution of the statistical properties of the correlation coefficients using the returns of the S\&P 500 stock market. We computed the mean, the absolute mean, the difference between absolute mean and mean, variance, skewness and kurtosis of the correlation coefficients $C_{ij}$, for short epochs of $M=20$ days and shifts of $\Delta \tau =1$ day and  $\Delta \tau =10$  days.  We also showed the evolution of  the mean of correlation coefficients, maximum eigenvalue of the correlation matrix, as well as the number of negative eigenvalues and smallest eigenvalue of the emerging spectrum, for the same epoch and shift.

Finally, we discussed the applications of RMT in financial markets. In an application, we demonstrated the use of RMT and correlation patterns in identifying possible ``market states'' and long-term precursors to the market crashes. In the second application, we presented the characterization of  catastrophic instabilities, i.e., the market crashes, using the smallest eigenvalue of the emerging spectra arising from correlation matrices computed over short epochs.

 \begin{acknowledgement}

The authors thank R. Chatterjee, S. Das and F. Leyvraz for the joint works presented here.
A.C. and K.S. acknowledge the support by grant number BT/BI/03/004/2003(C) of Govt. of India, Ministry of Science and Technology, Department of Biotechnology, Bioinformatics division, University of Potential Excellence-II grant (Project ID-47) of JNU, New Delhi, and the DST-PURSE grant given to JNU by the Department of Science and Technology, Government of India.
K.S. acknowledges the University Grants Commission (Ministry of Human Research Development, Govt. of India) for her senior research fellowship. H.K.P. is grateful for postdoctoral fellowship provided by UNAM-DGAPA. A.C., H.K.P., K.S. and T.H.S. acknowledge support by Project CONACyT Fronteras 201, and also support from the project UNAM-DGAPA-PAPIIT IG 100616. 
\end{acknowledgement}

%>>>>>>>>>>>>>>>>>>>>>>>>>>>>>>>>>>>>>>>>>>>>>>>>>>>>>>>>>>>>>>>>>>>>>>>>>>>>>>>>>>>>>>>>>>
\bibliographystyle{spmpsci}
\bibliography{Book_chapter_2018}
\end{document}